\documentclass[a4paper]{article}

\usepackage{graphicx}
\usepackage[hmargin=0cm,vmargin=0cm]{geometry}

\begin{document}

\begin{figure}
\includegraphics{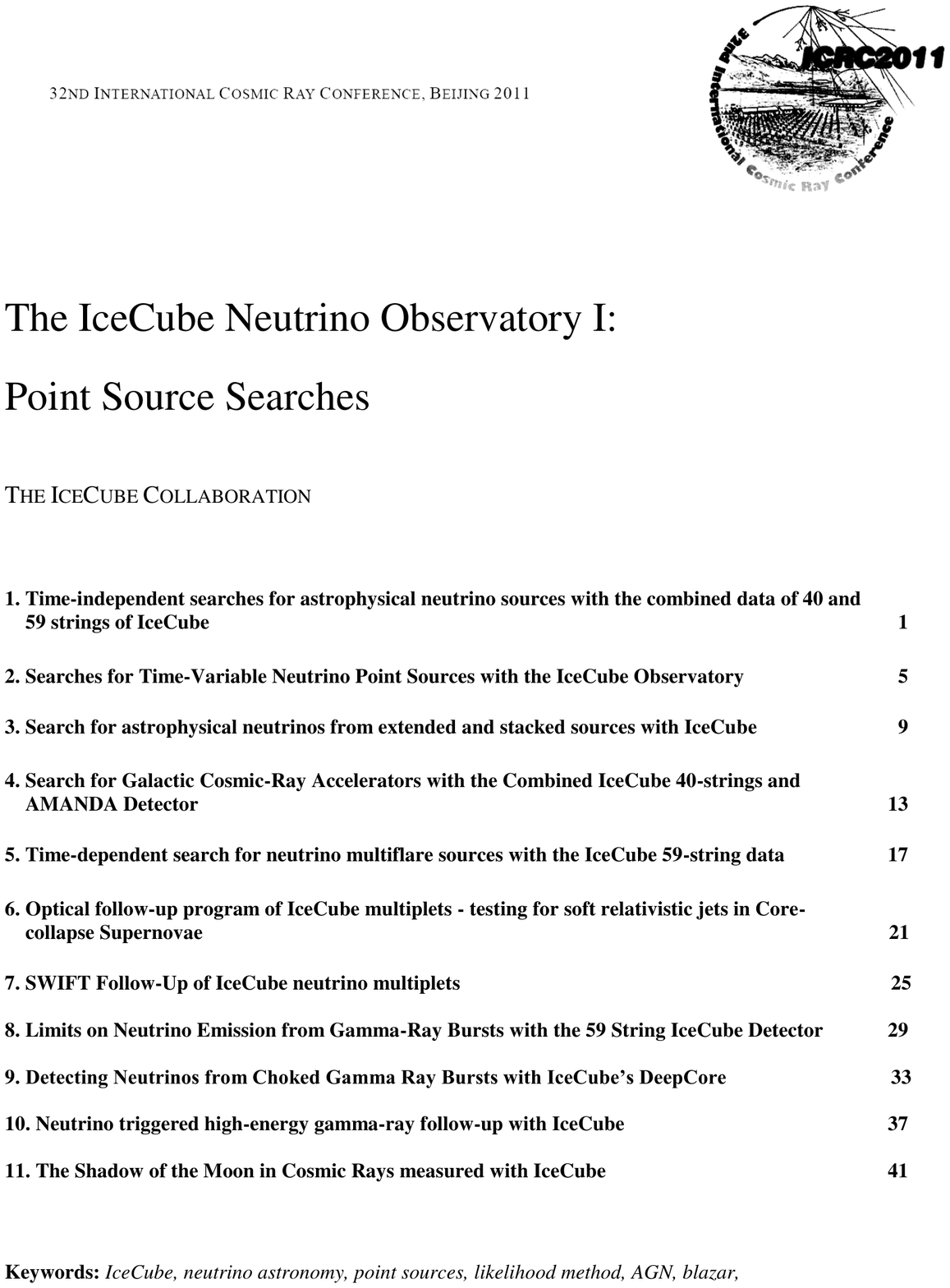}
\end{figure}
\clearpage

\begin{figure}
\includegraphics{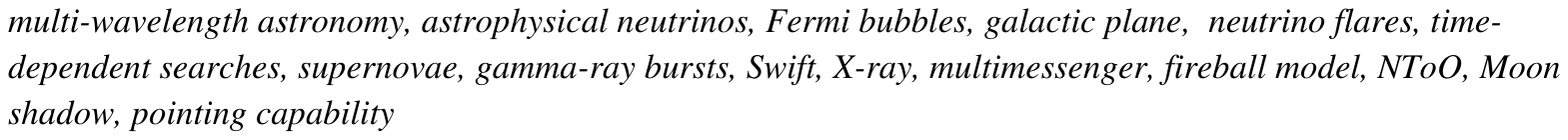}
\end{figure}
\clearpage

\begin{figure}
\includegraphics{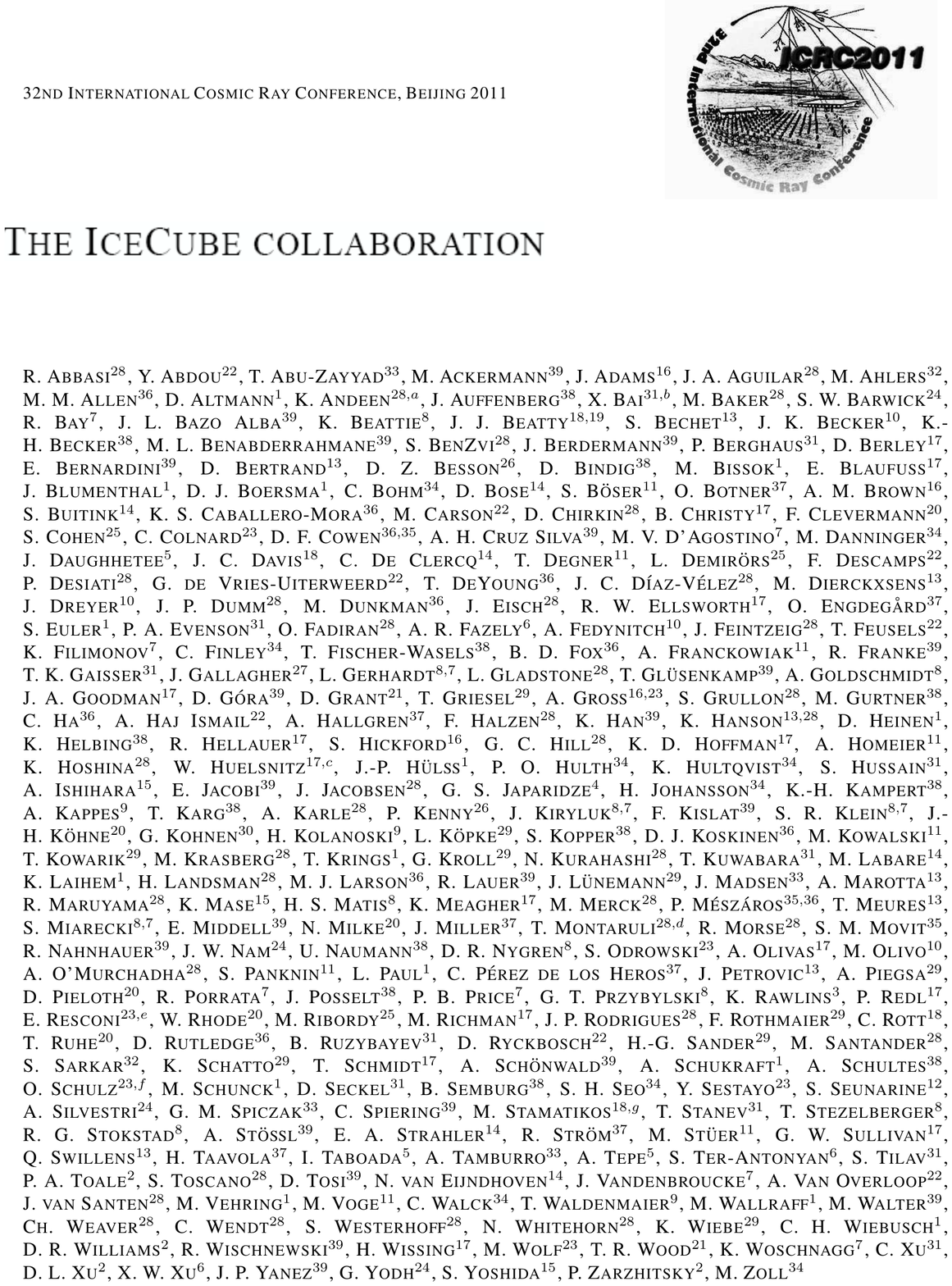}
\end{figure}
\clearpage

\begin{figure}
\includegraphics{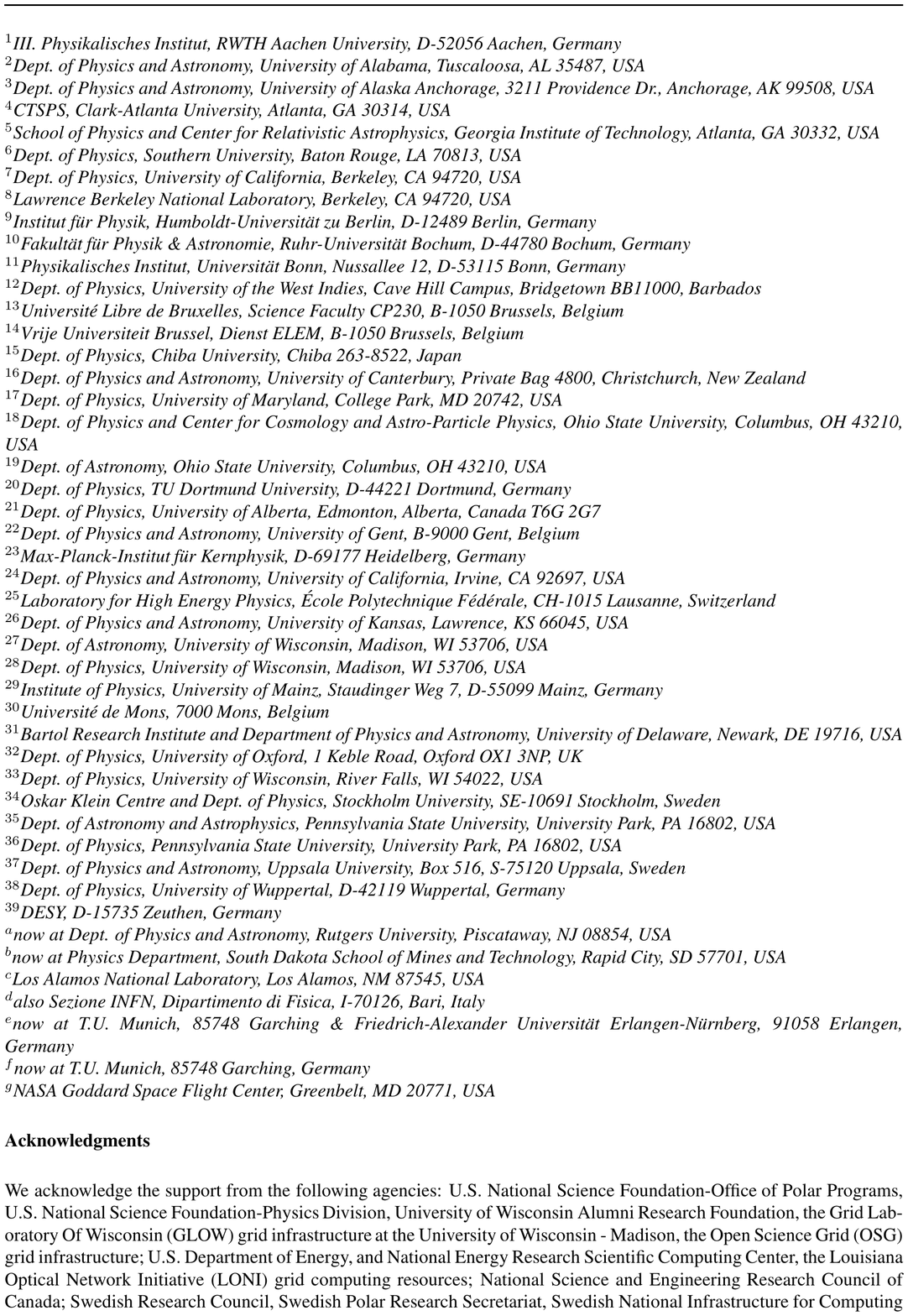}
\end{figure}
\clearpage

\begin{figure}
\includegraphics{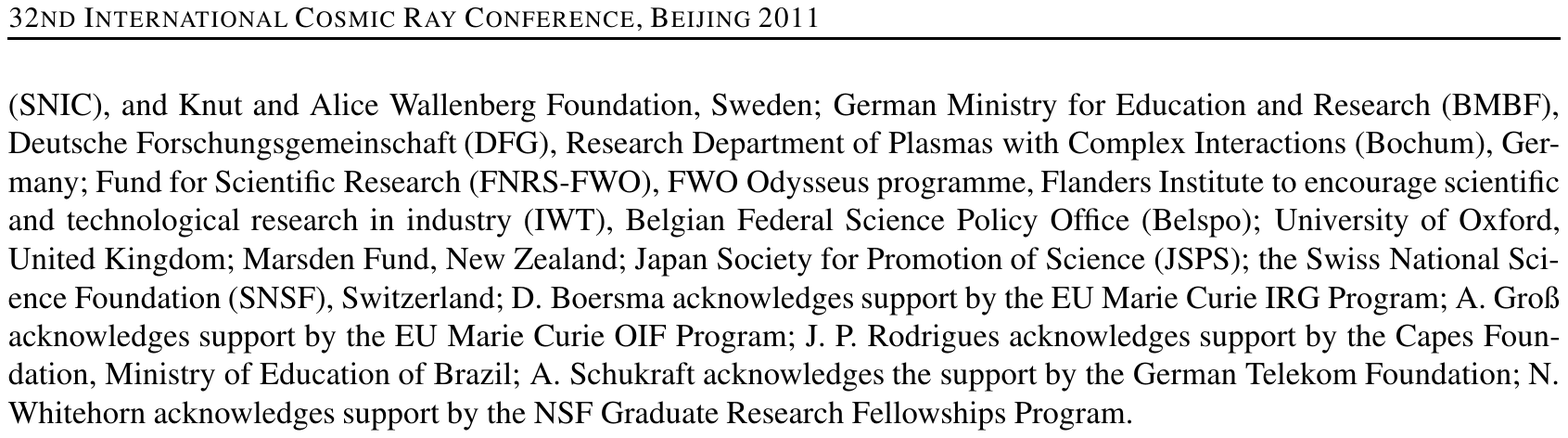}
\end{figure}
\clearpage

\begin{figure}
\includegraphics{6.pdf}
\end{figure}
\clearpage

\begin{figure}
\includegraphics{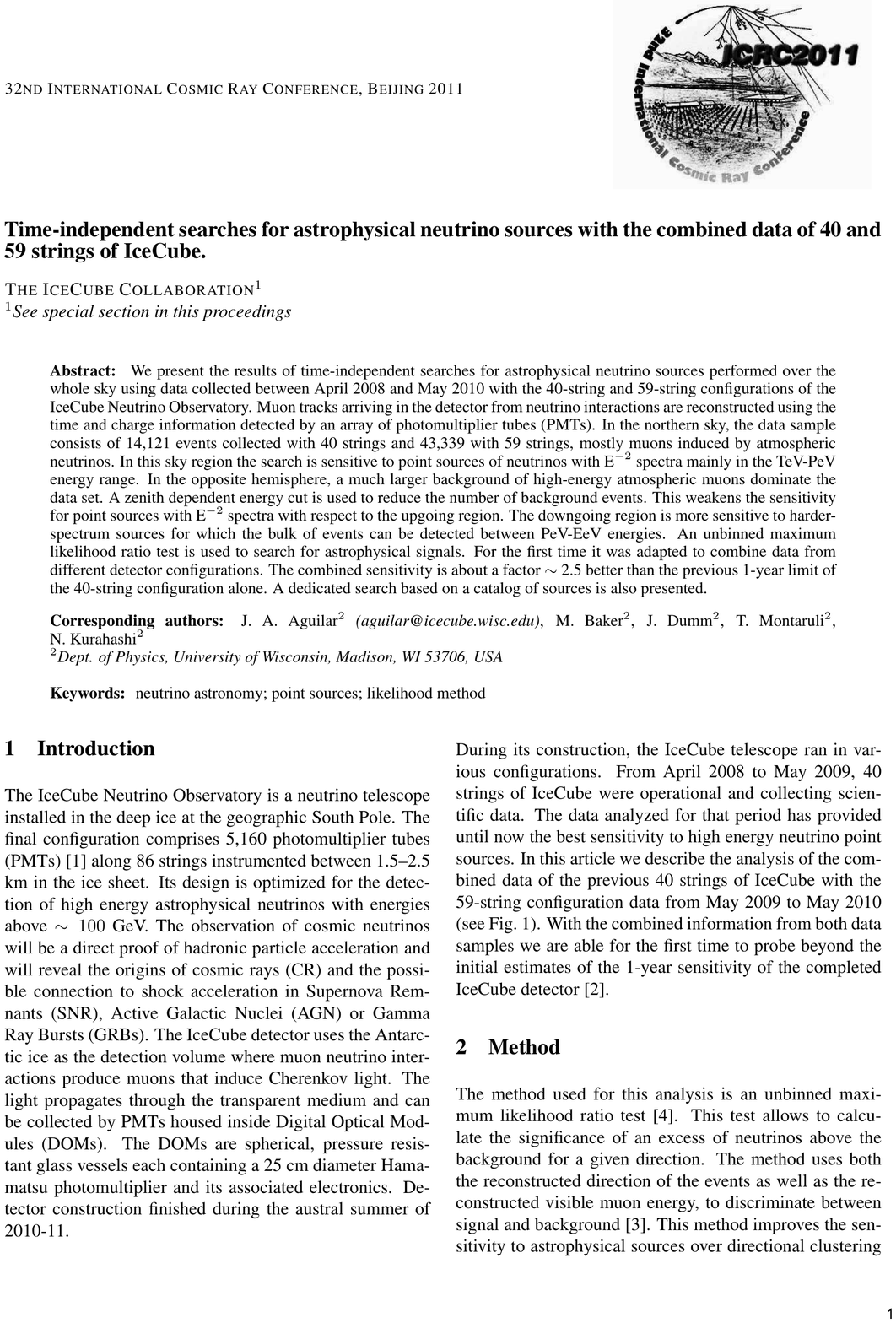}
\end{figure}
\clearpage

\begin{figure}
\includegraphics{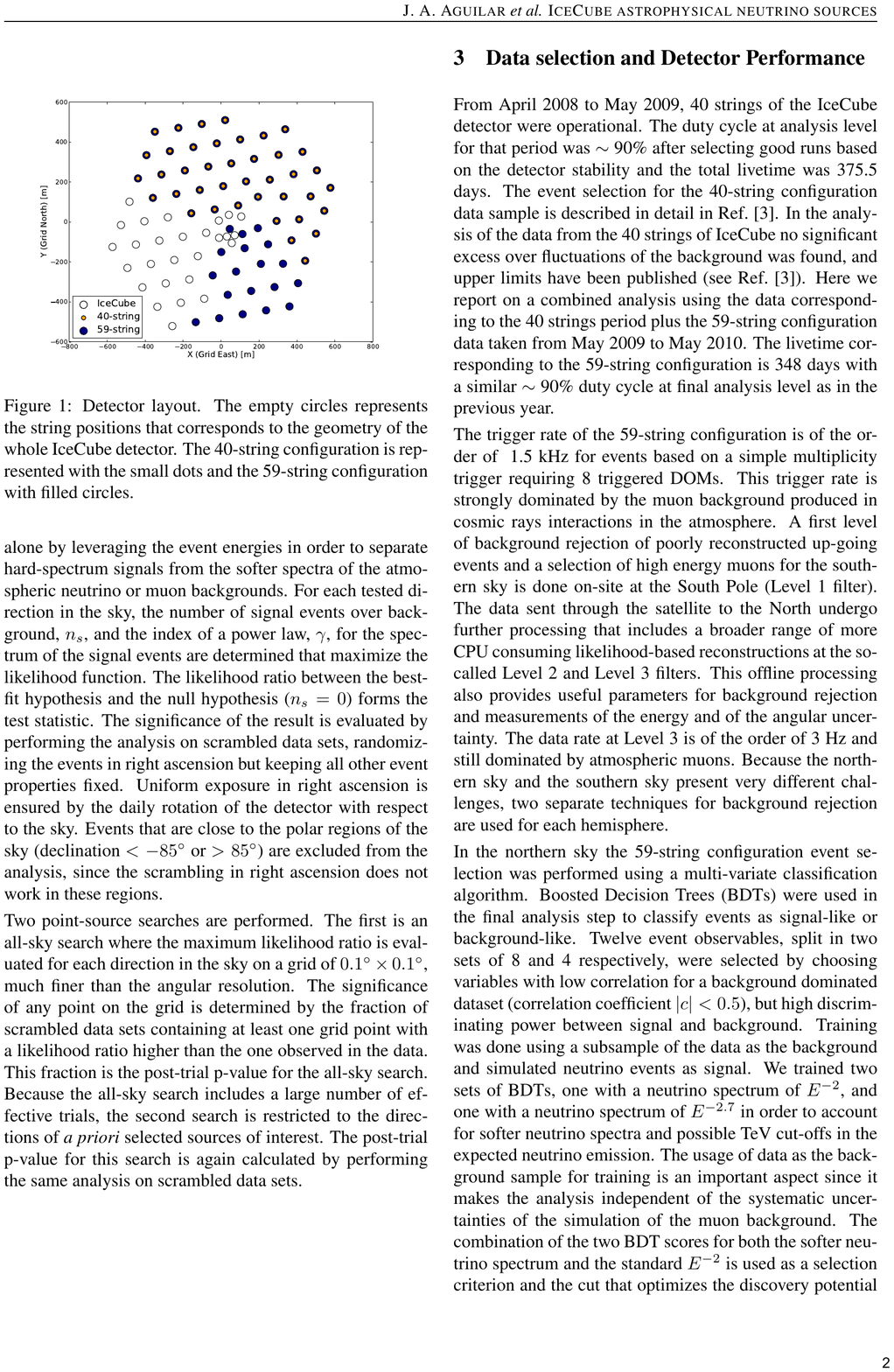}
\end{figure}
\clearpage

\begin{figure}
\includegraphics{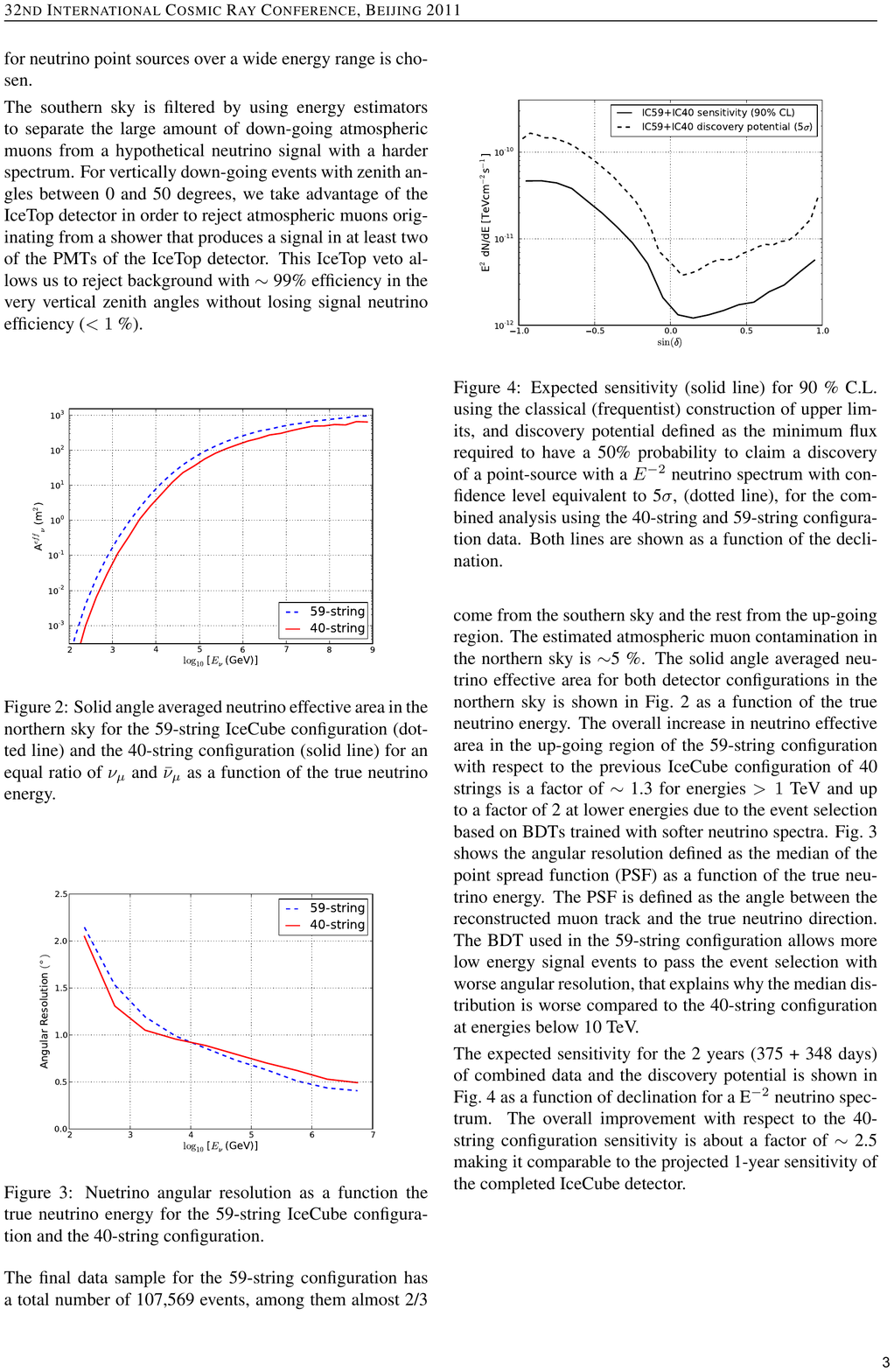}
\end{figure}
\clearpage

\begin{figure}
\includegraphics{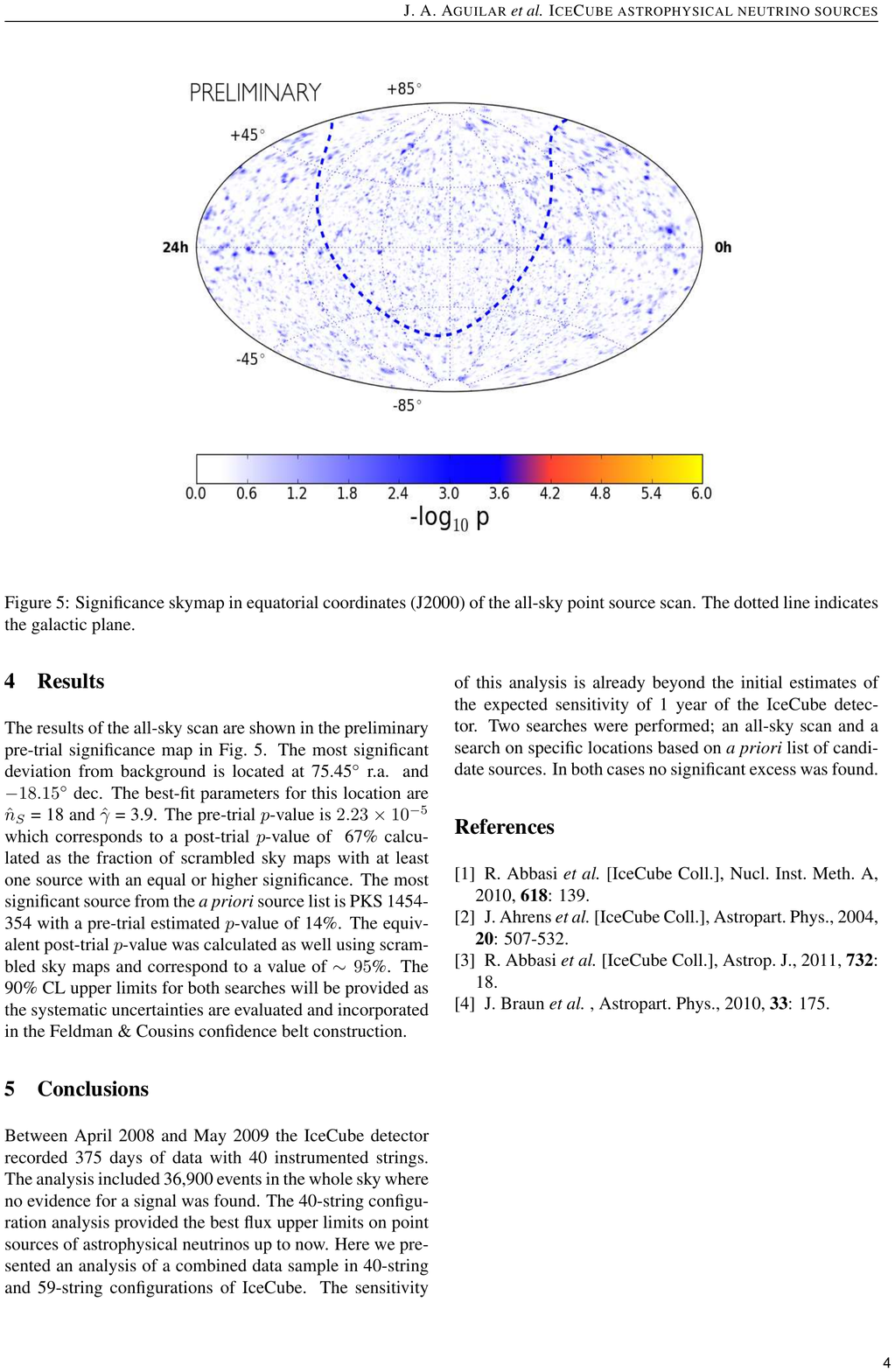}
\end{figure}
\clearpage

\begin{figure}
\includegraphics{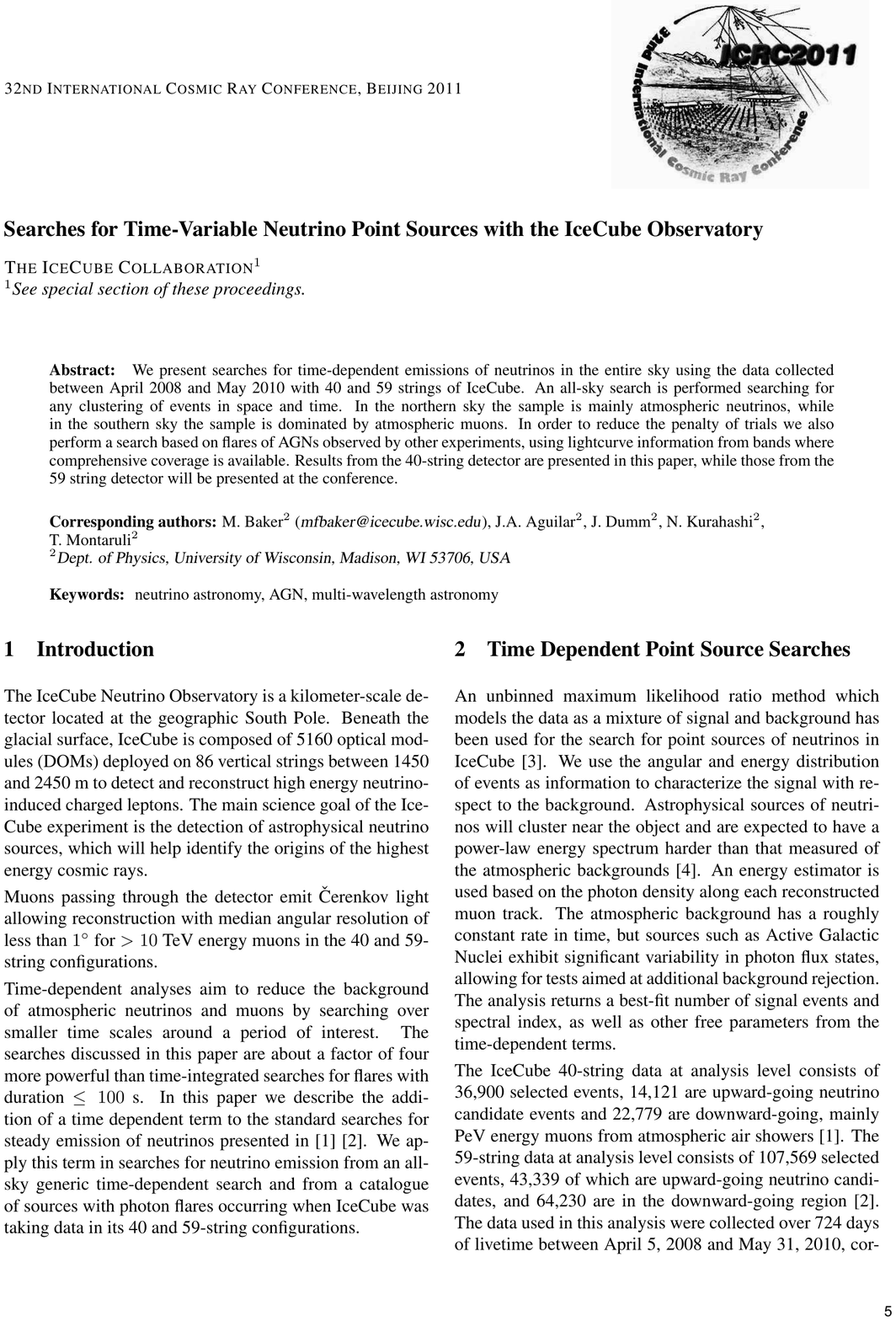}
\end{figure}
\clearpage

\begin{figure}
\includegraphics{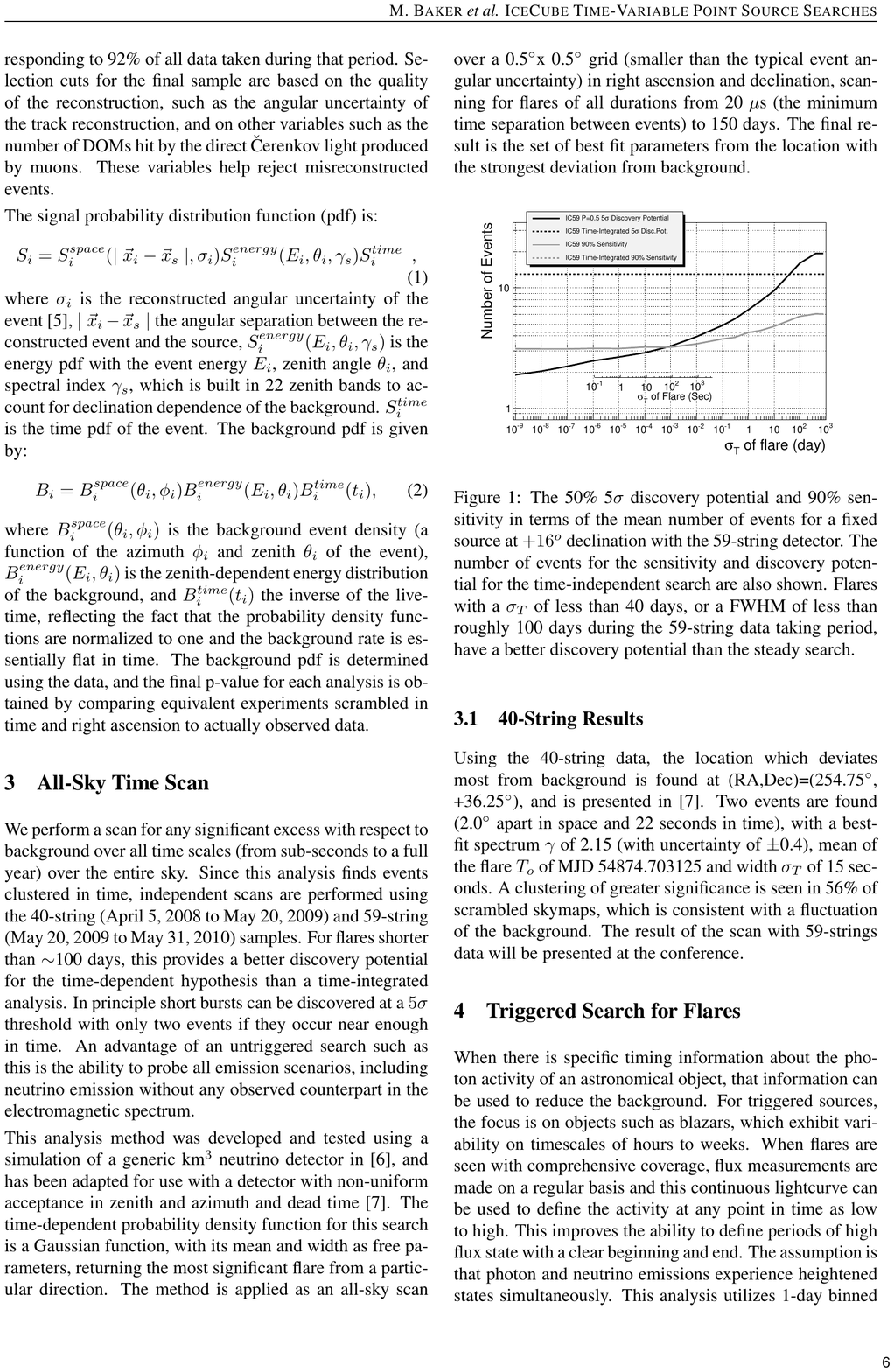}
\end{figure}
\clearpage

\begin{figure}
\includegraphics{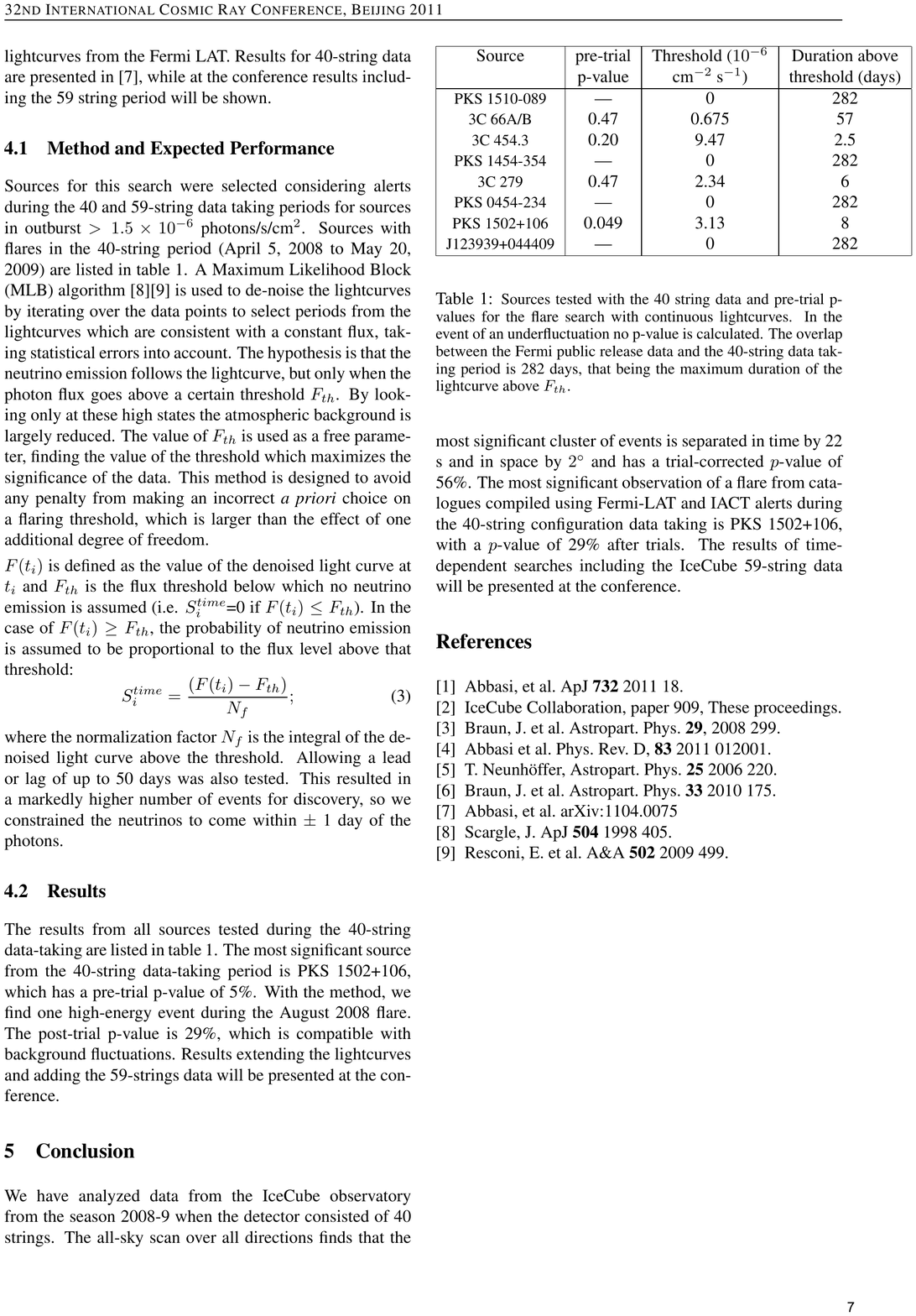}
\end{figure}
\clearpage

\begin{figure}
\includegraphics{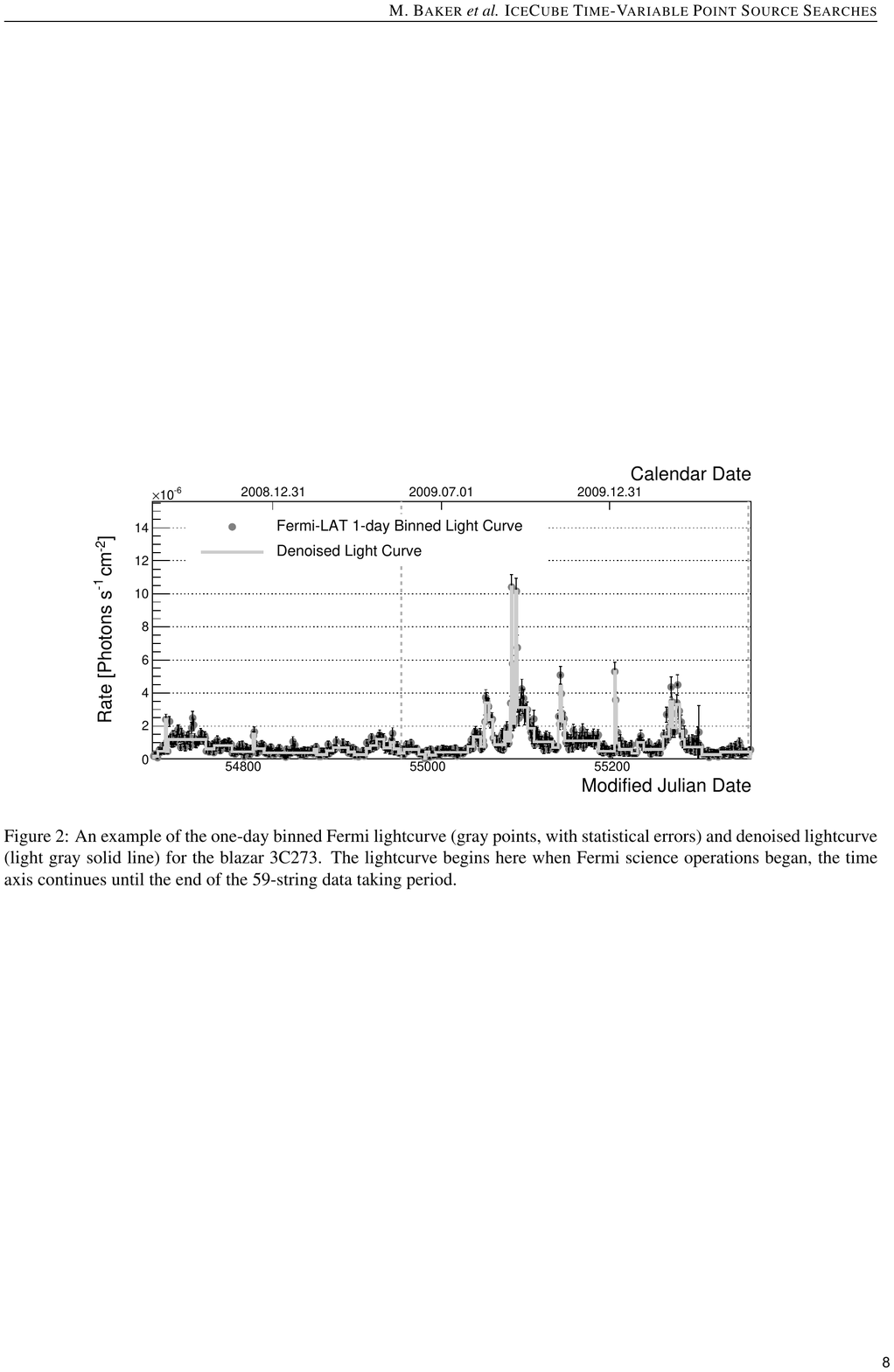}
\end{figure}
\clearpage

\begin{figure}
\includegraphics{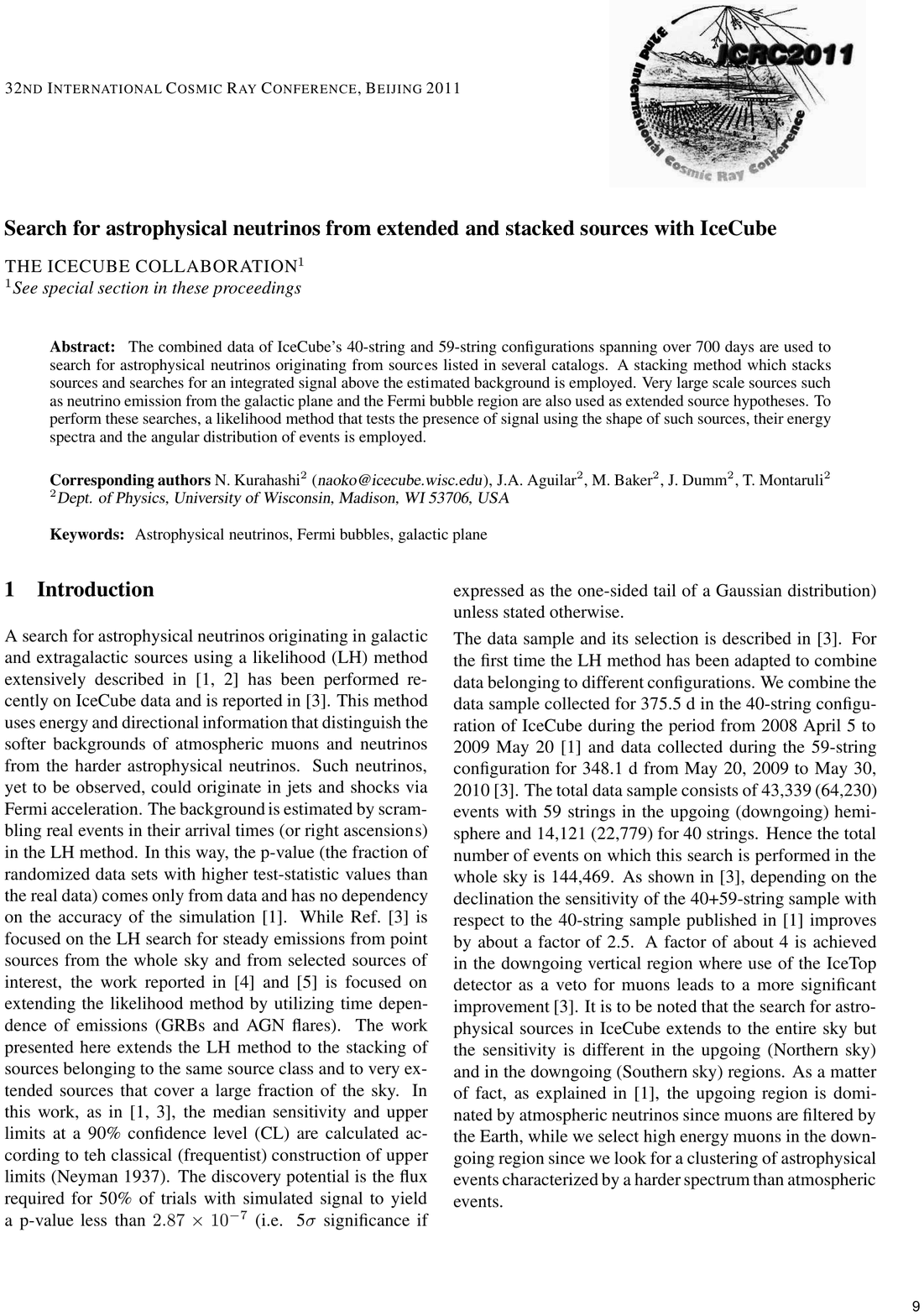}
\end{figure}
\clearpage

\begin{figure}
\includegraphics{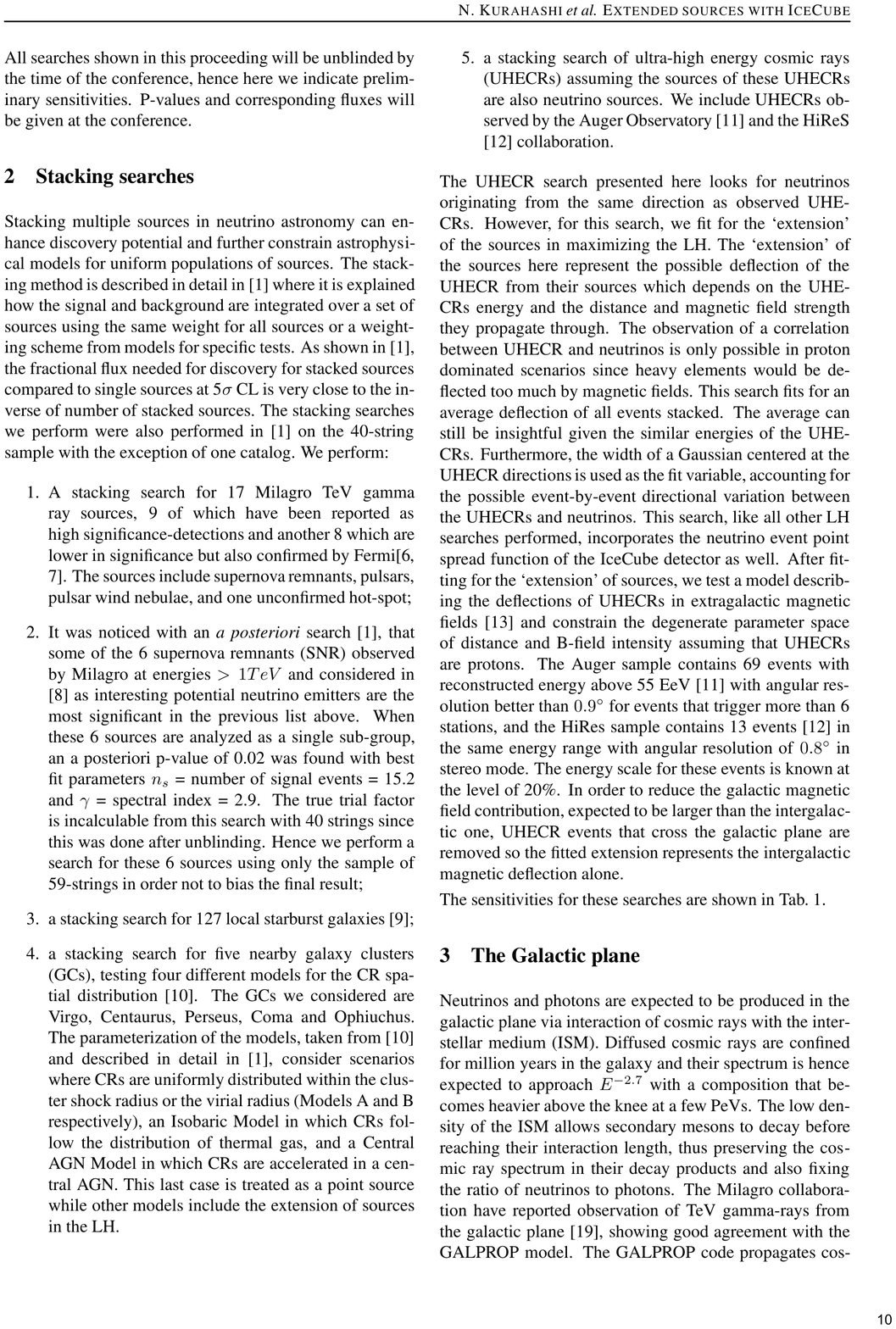}
\end{figure}
\clearpage

\begin{figure}
\includegraphics{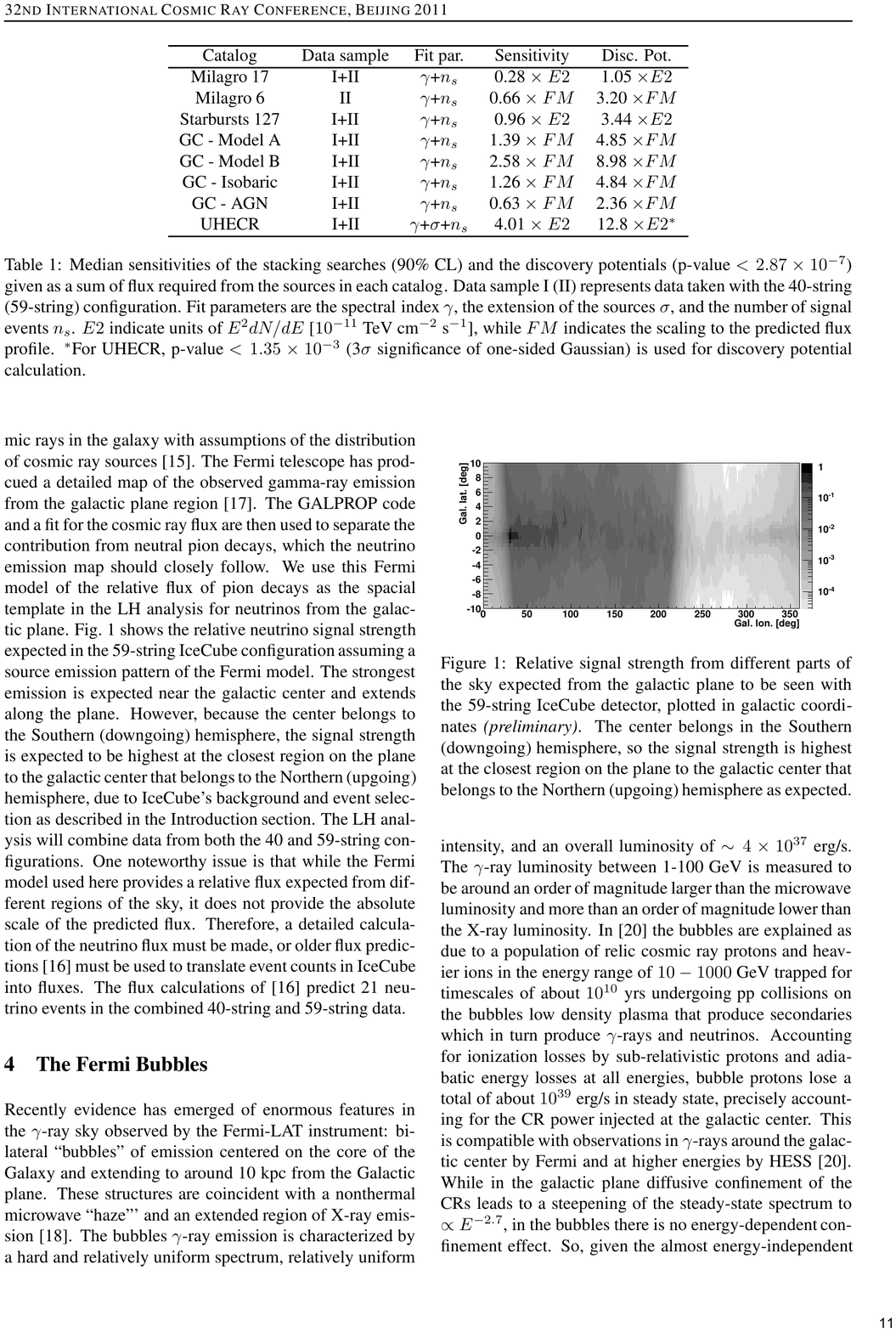}
\end{figure}
\clearpage

\begin{figure}
\includegraphics{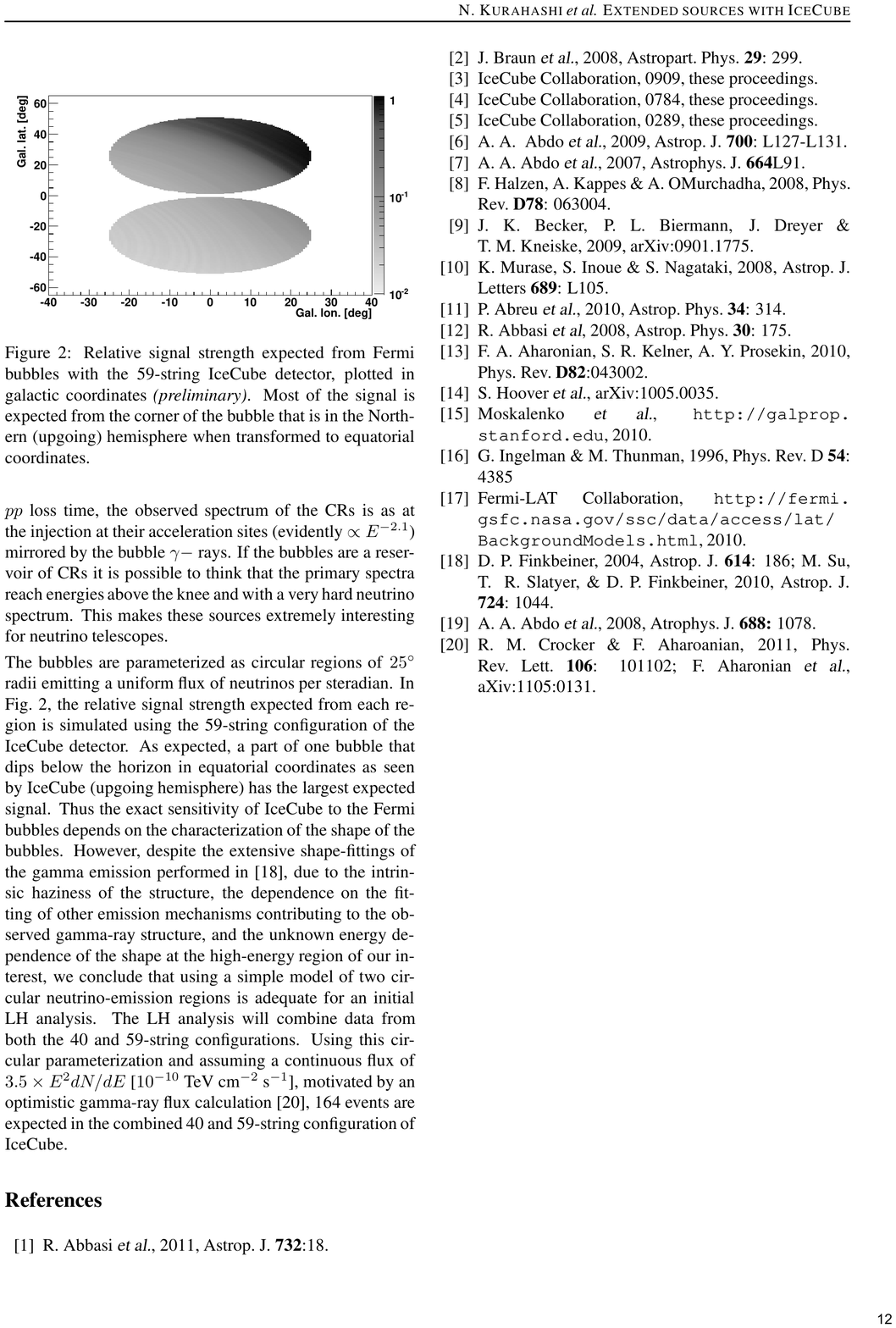}
\end{figure}
\clearpage

\begin{figure}
\includegraphics{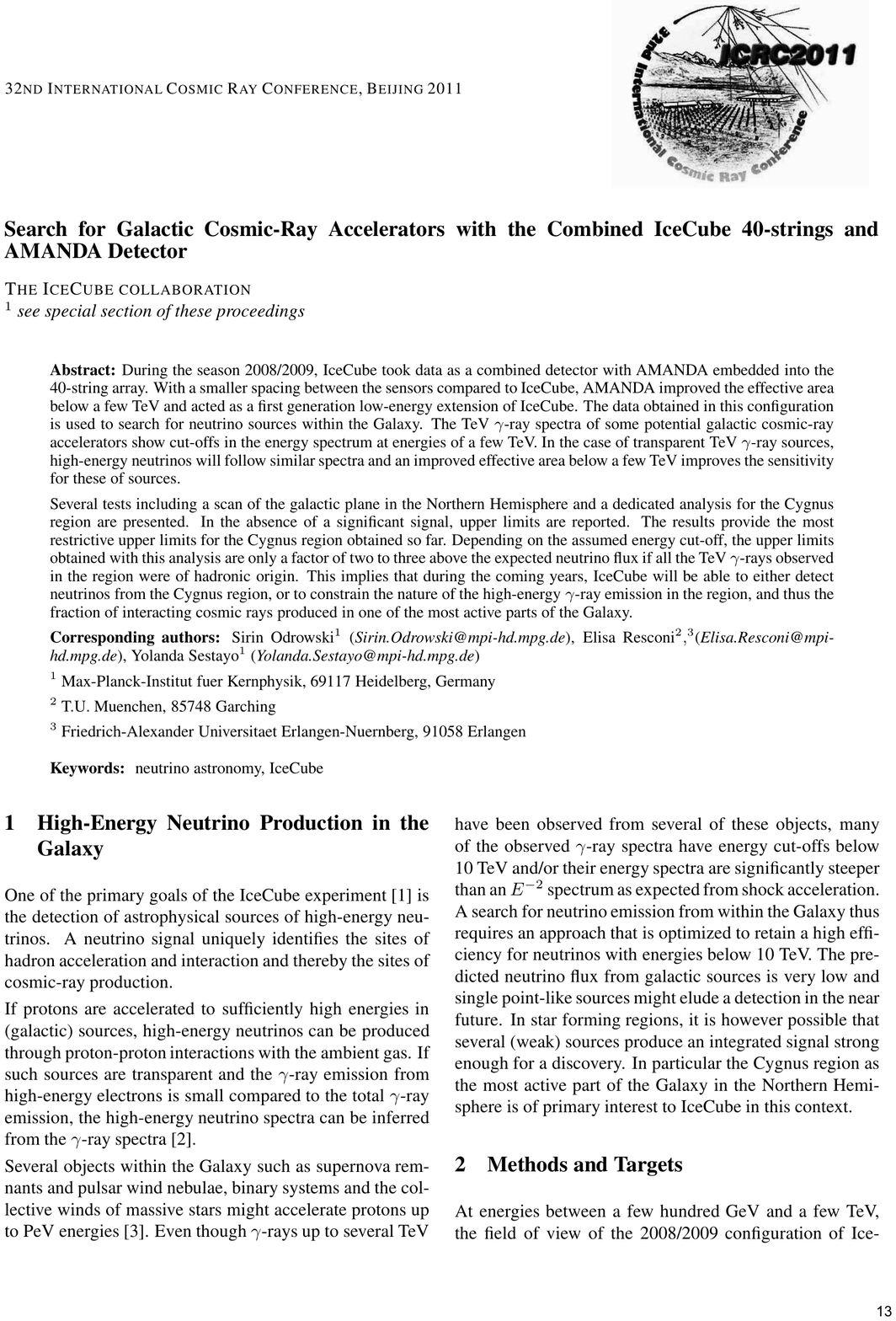}
\end{figure}
\clearpage

\begin{figure}
\includegraphics{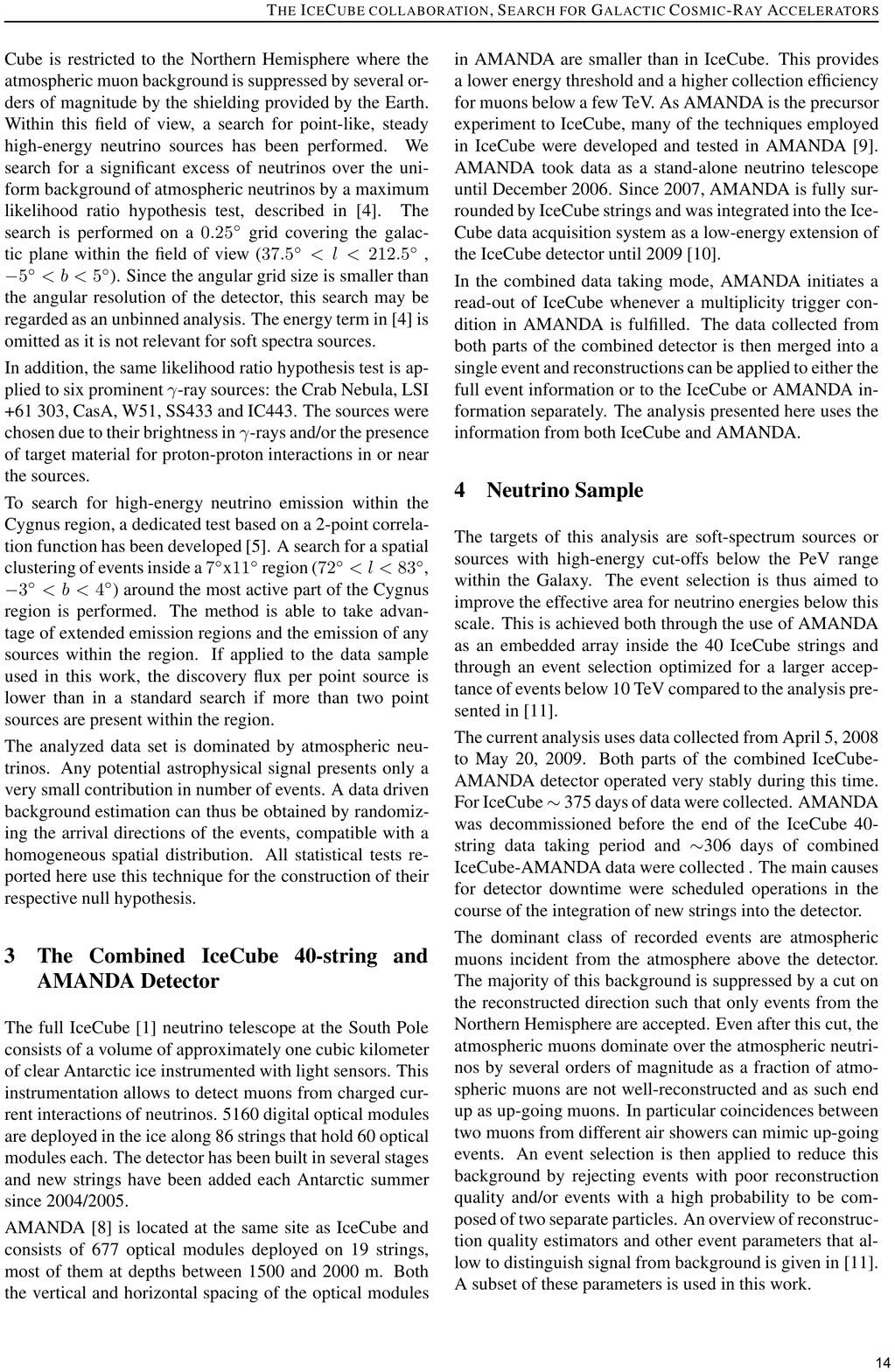}
\end{figure}
\clearpage

\begin{figure}
\includegraphics{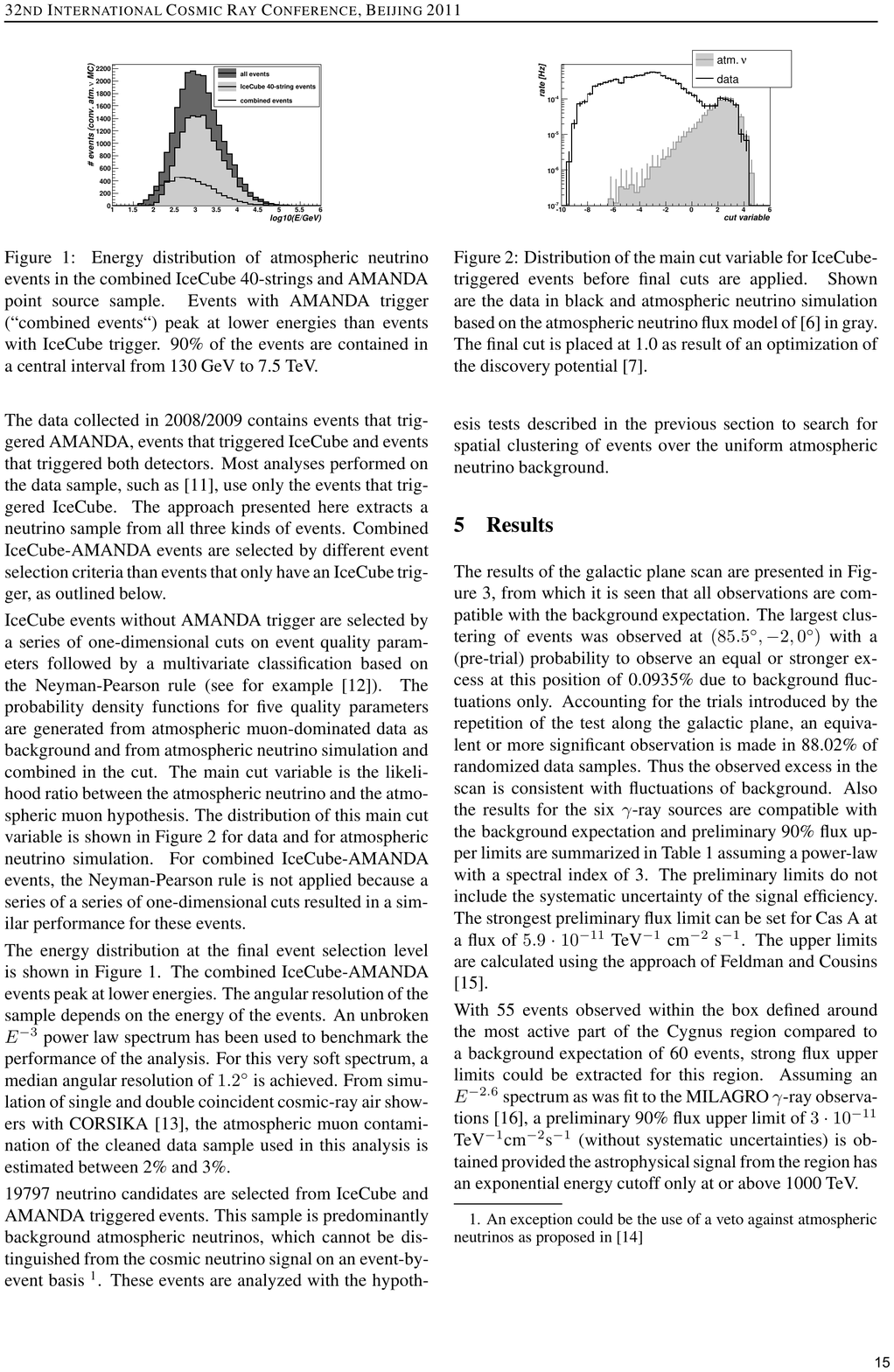}
\end{figure}
\clearpage

\begin{figure}
\includegraphics{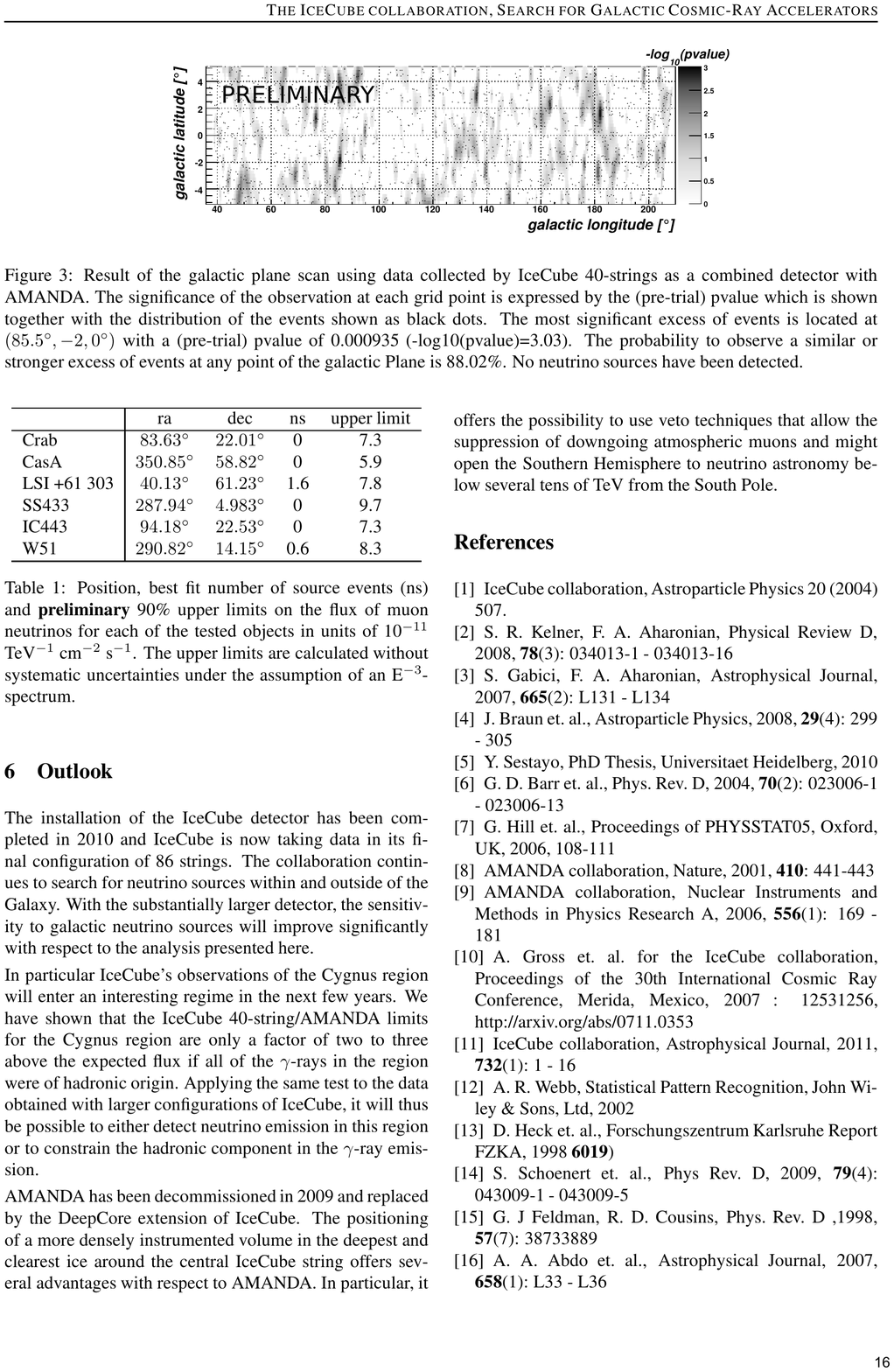}
\end{figure}
\clearpage

\begin{figure}
\includegraphics{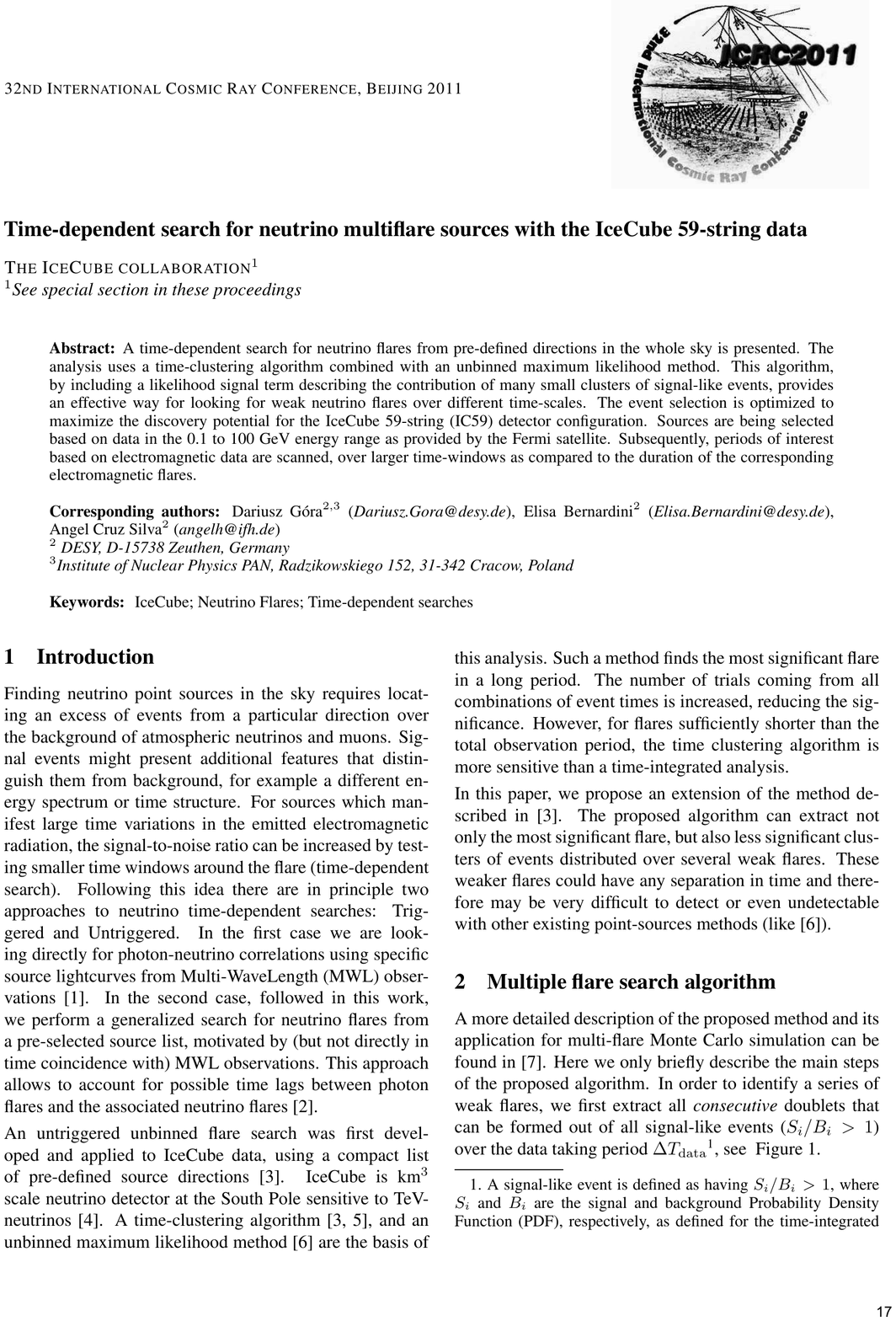}
\end{figure}
\clearpage

\begin{figure}
\includegraphics{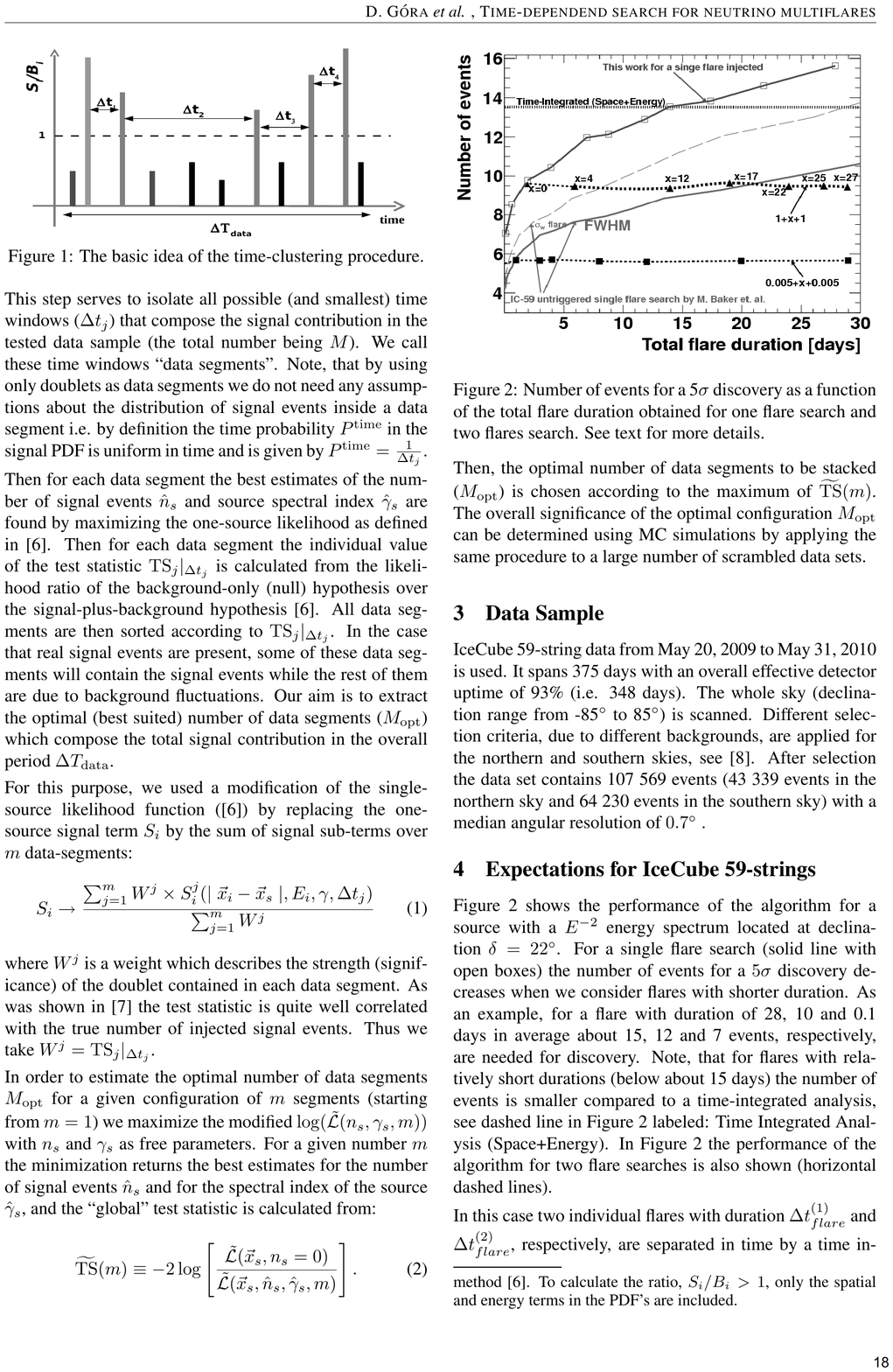}
\end{figure}
\clearpage

\begin{figure}
\includegraphics{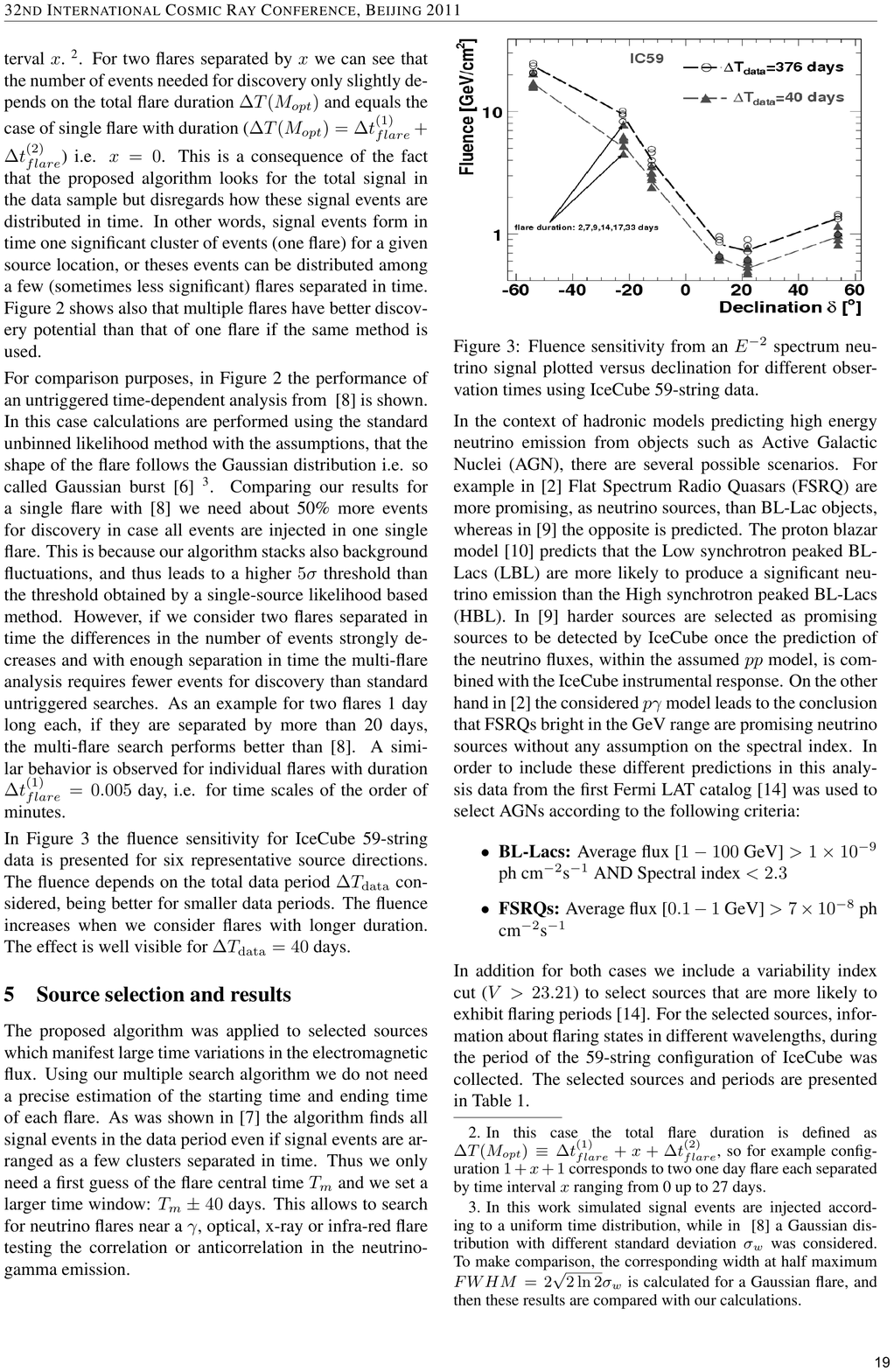}
\end{figure}
\clearpage

\begin{figure}
\includegraphics{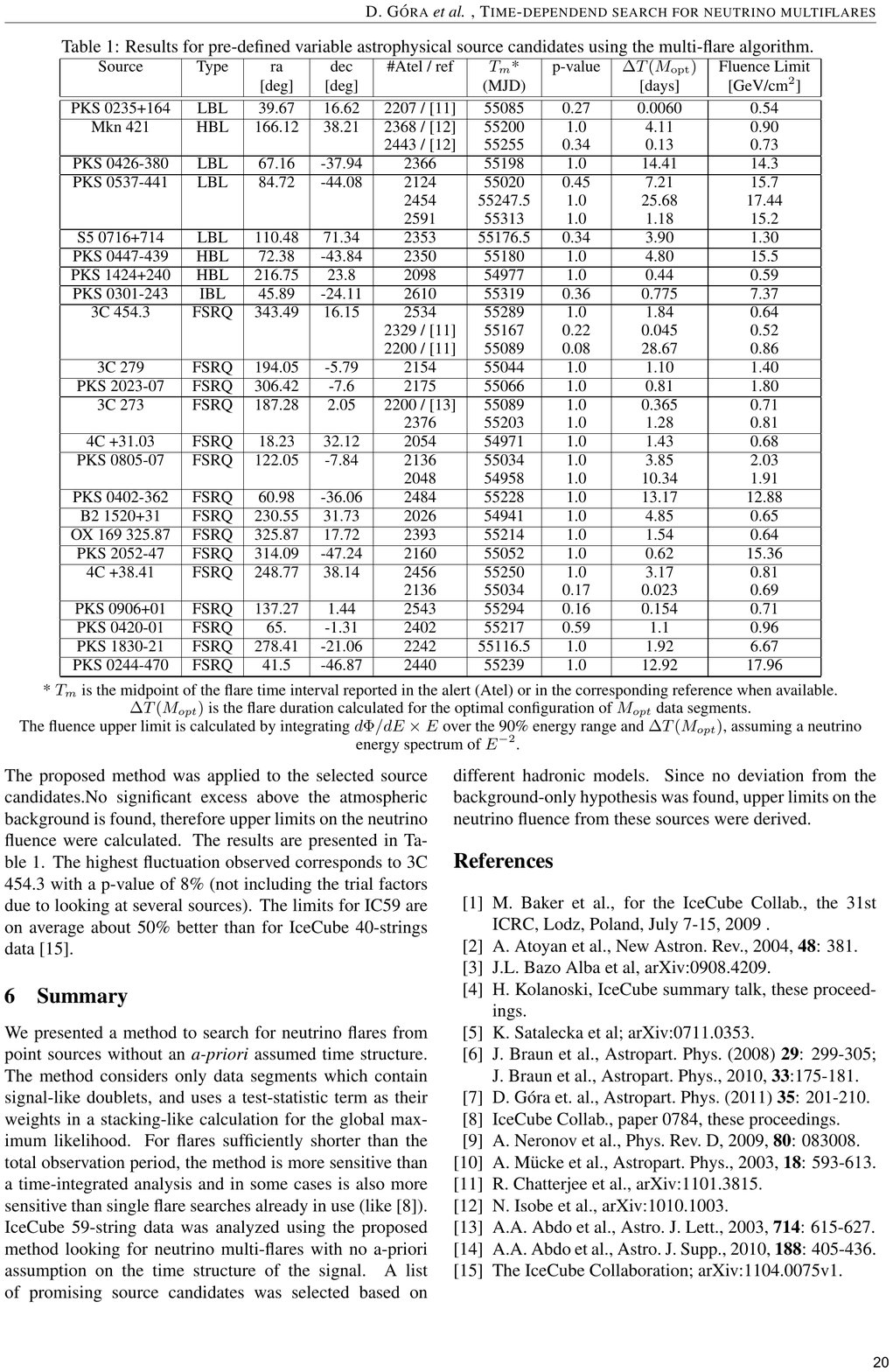}
\end{figure}
\clearpage

\begin{figure}
\includegraphics{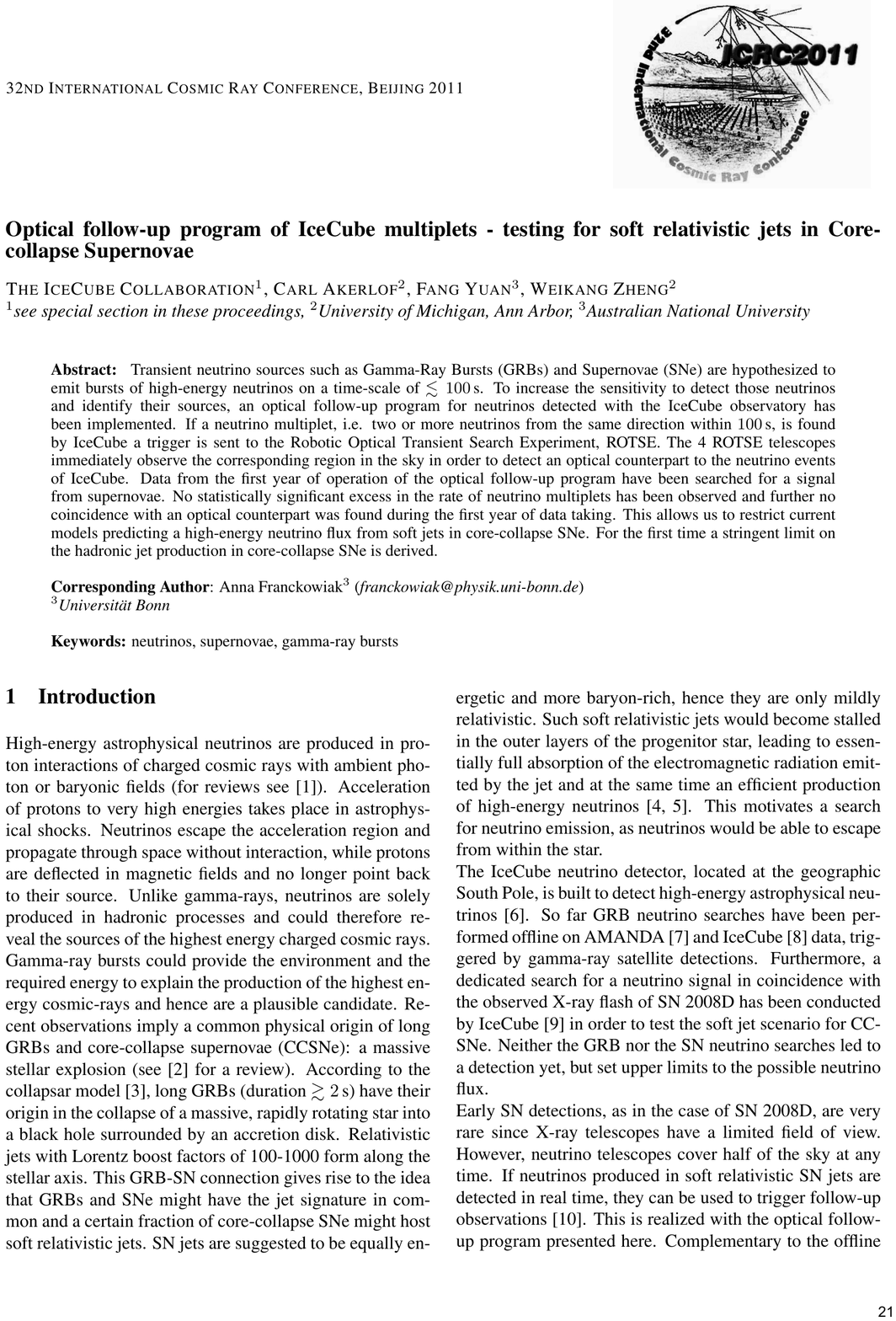}
\end{figure}
\clearpage

\begin{figure}
\includegraphics{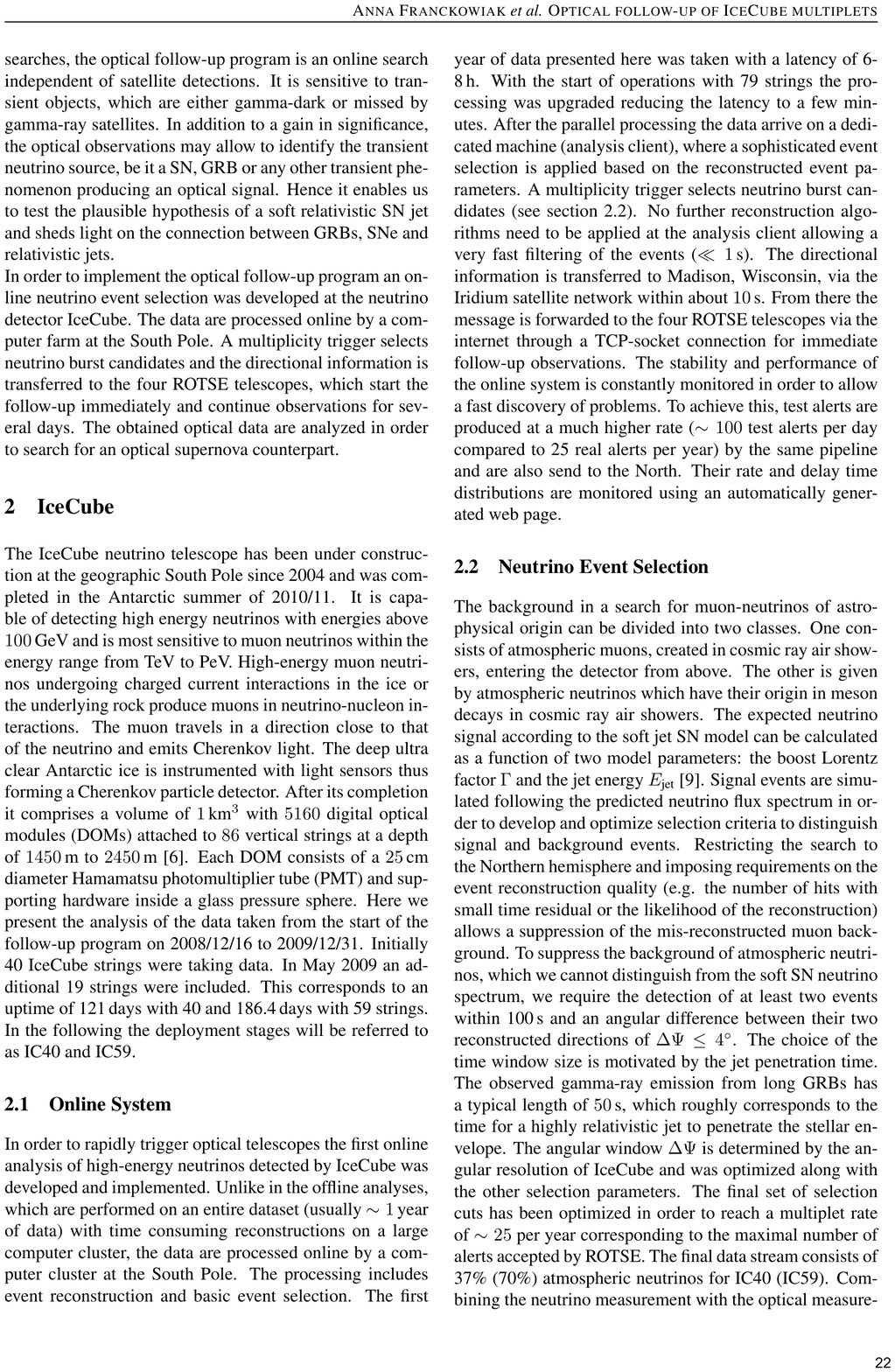}
\end{figure}
\clearpage

\begin{figure}
\includegraphics{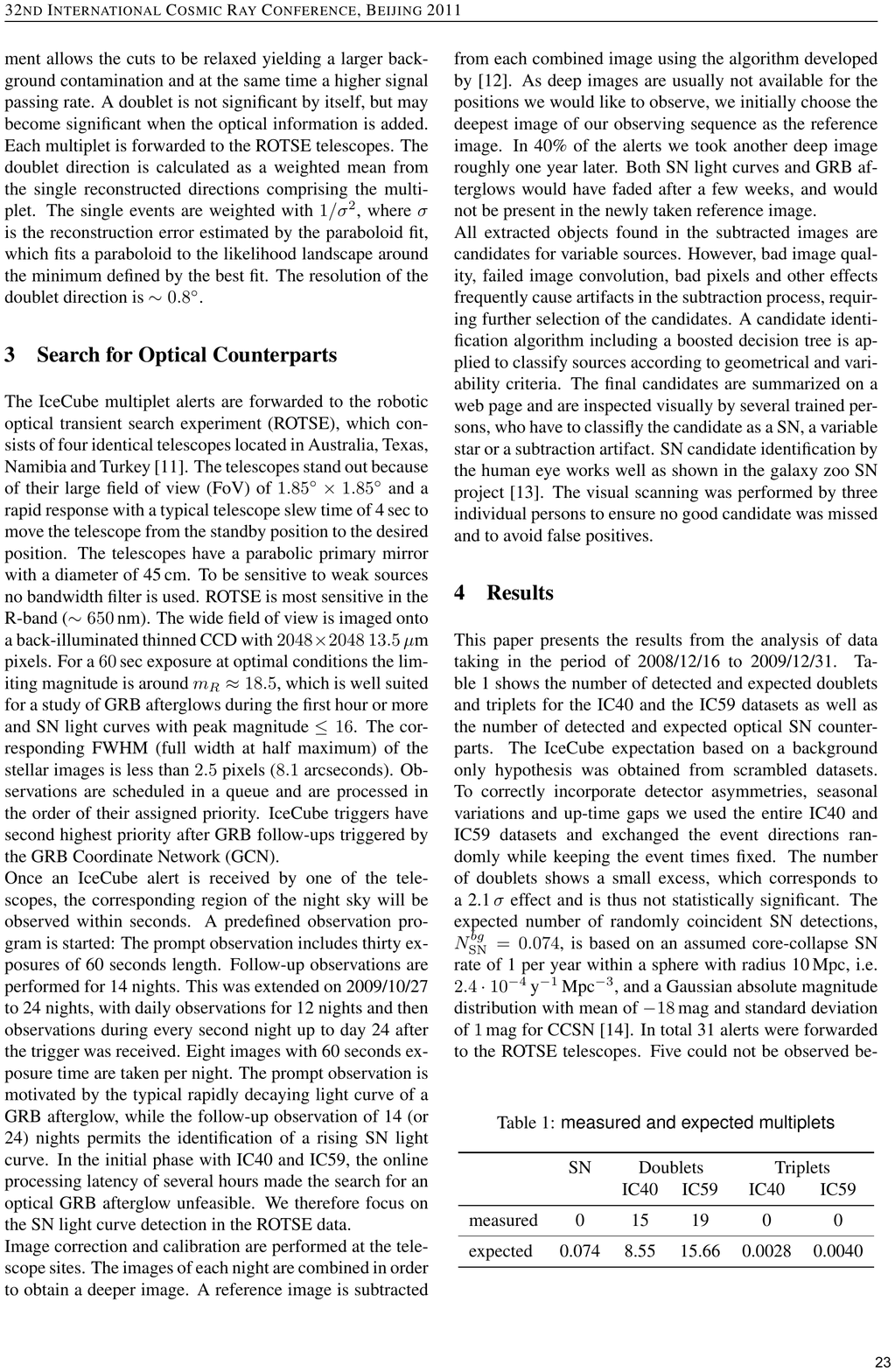}
\end{figure}
\clearpage

\begin{figure}
\includegraphics{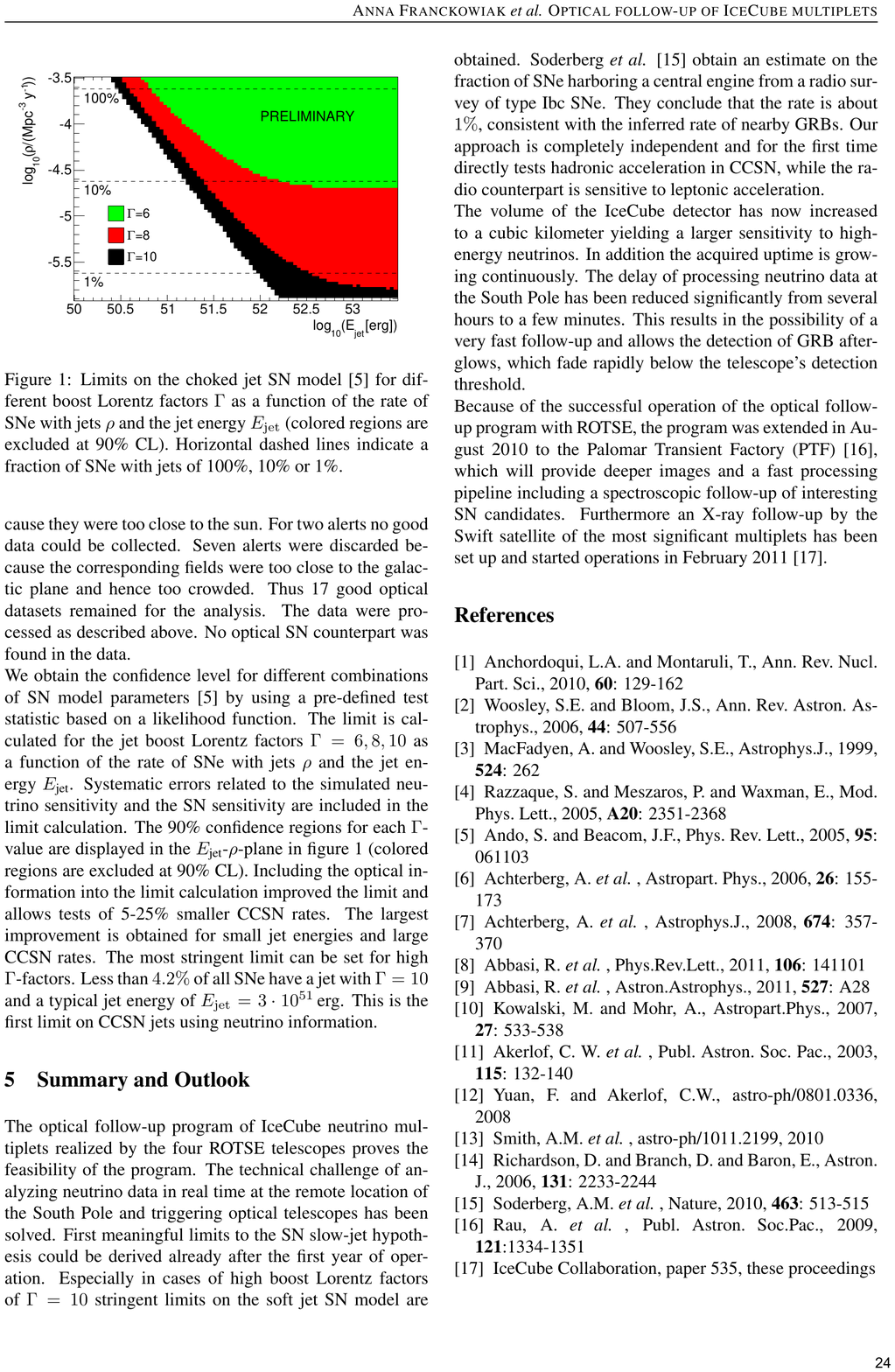}
\end{figure}
\clearpage

\begin{figure}
\includegraphics{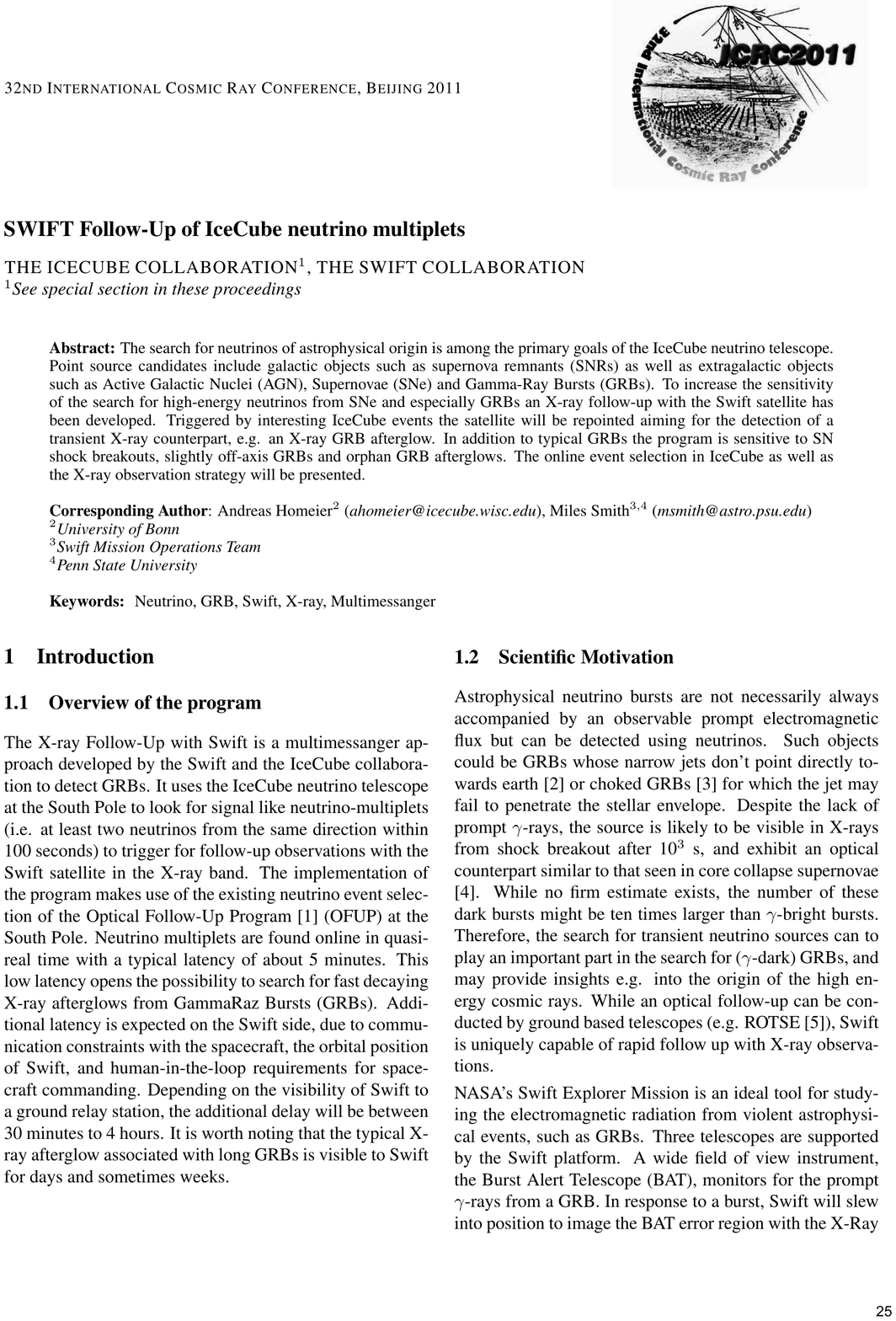}
\end{figure}
\clearpage

\begin{figure}
\includegraphics{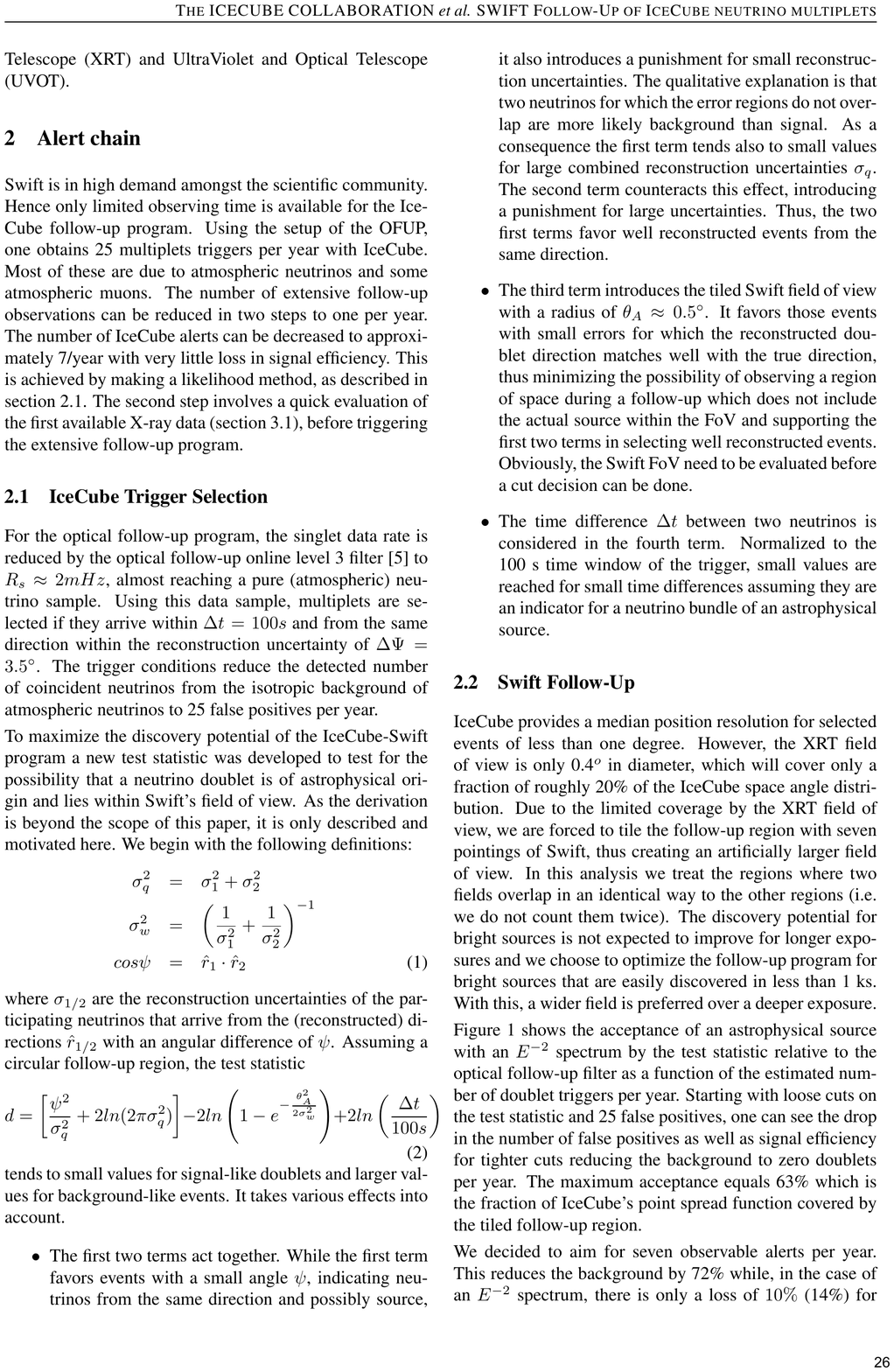}
\end{figure}
\clearpage

\begin{figure}
\includegraphics{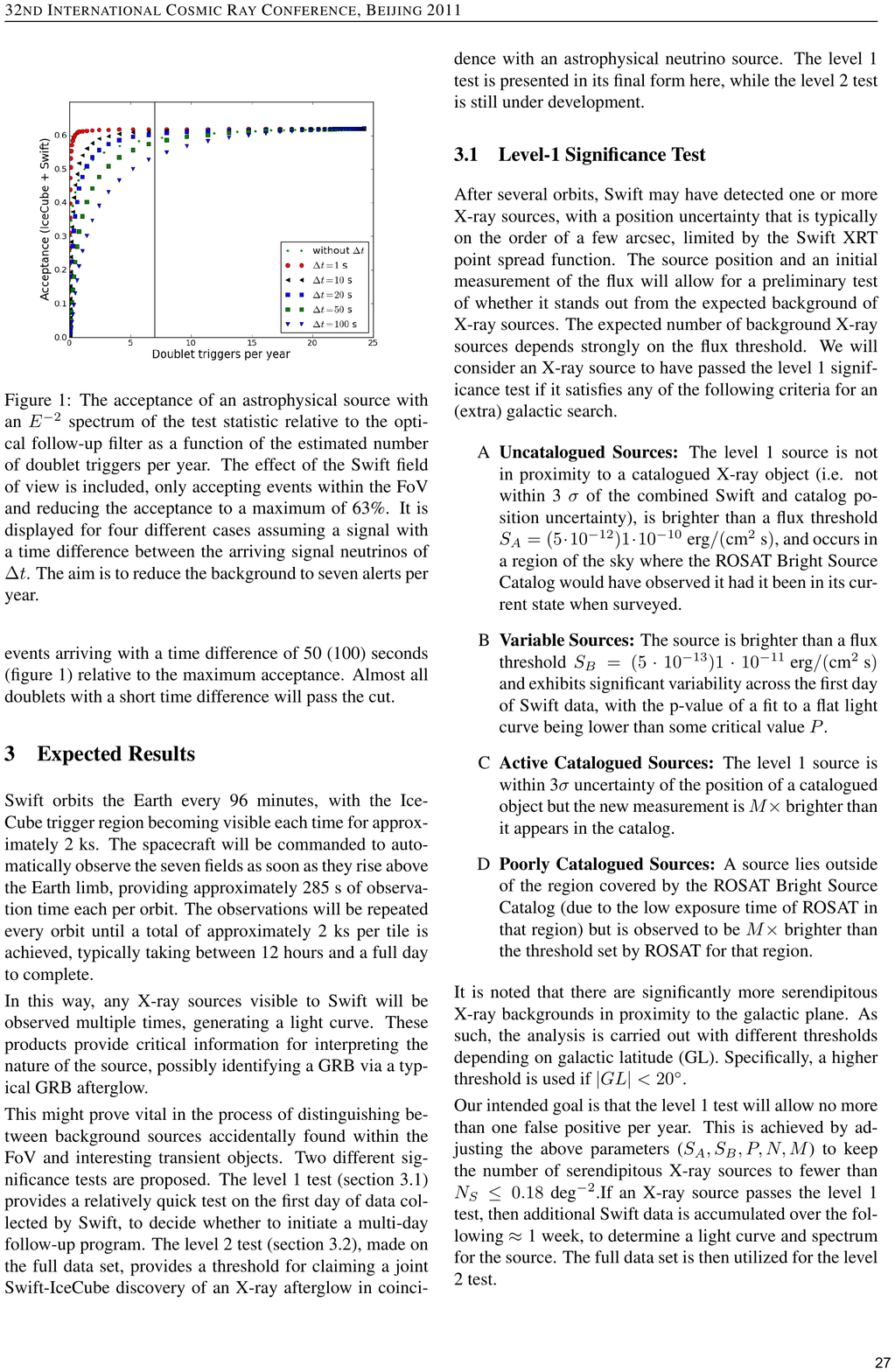}
\end{figure}
\clearpage

\begin{figure}
\includegraphics{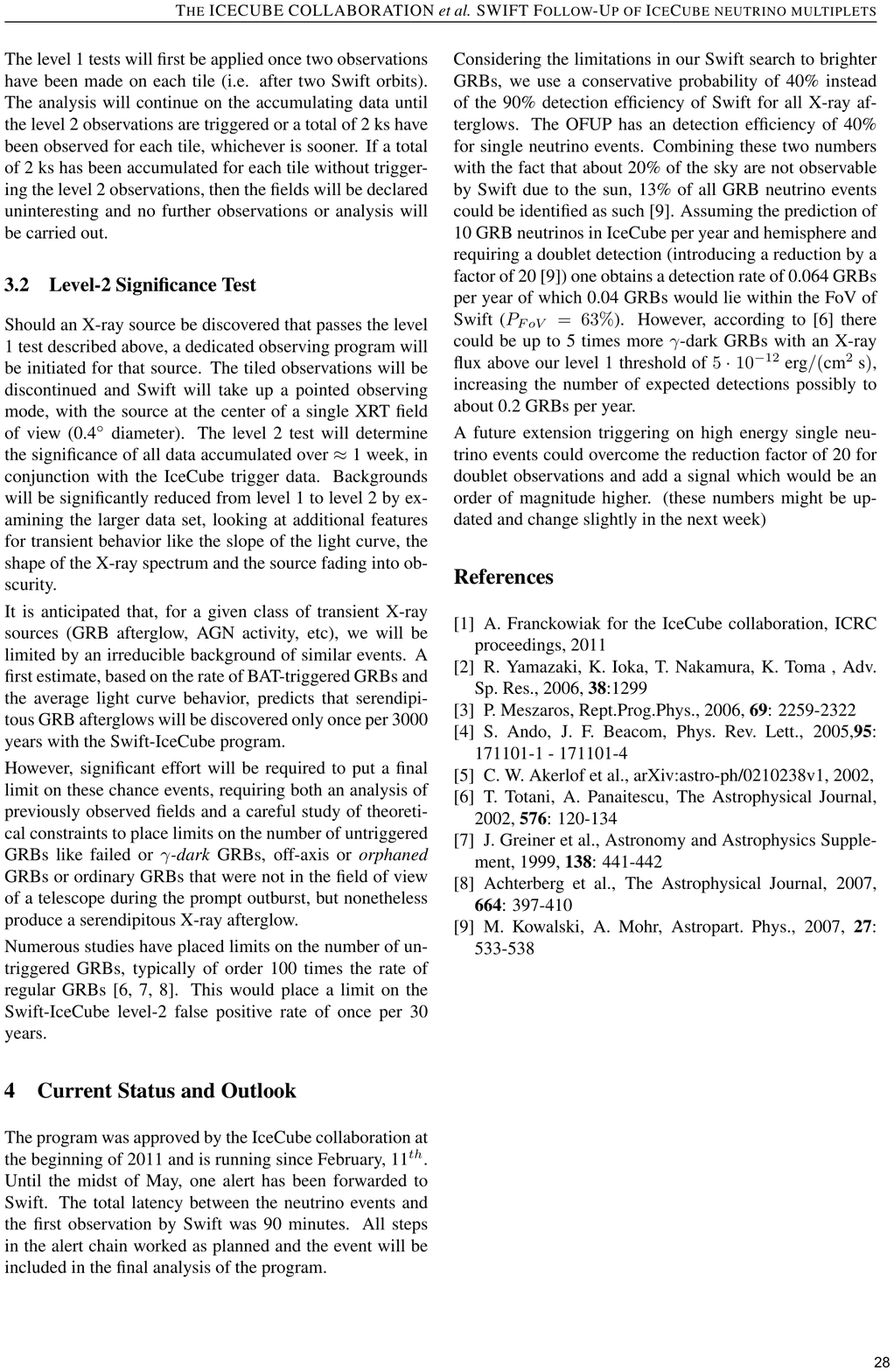}
\end{figure}
\clearpage

\begin{figure}
\includegraphics{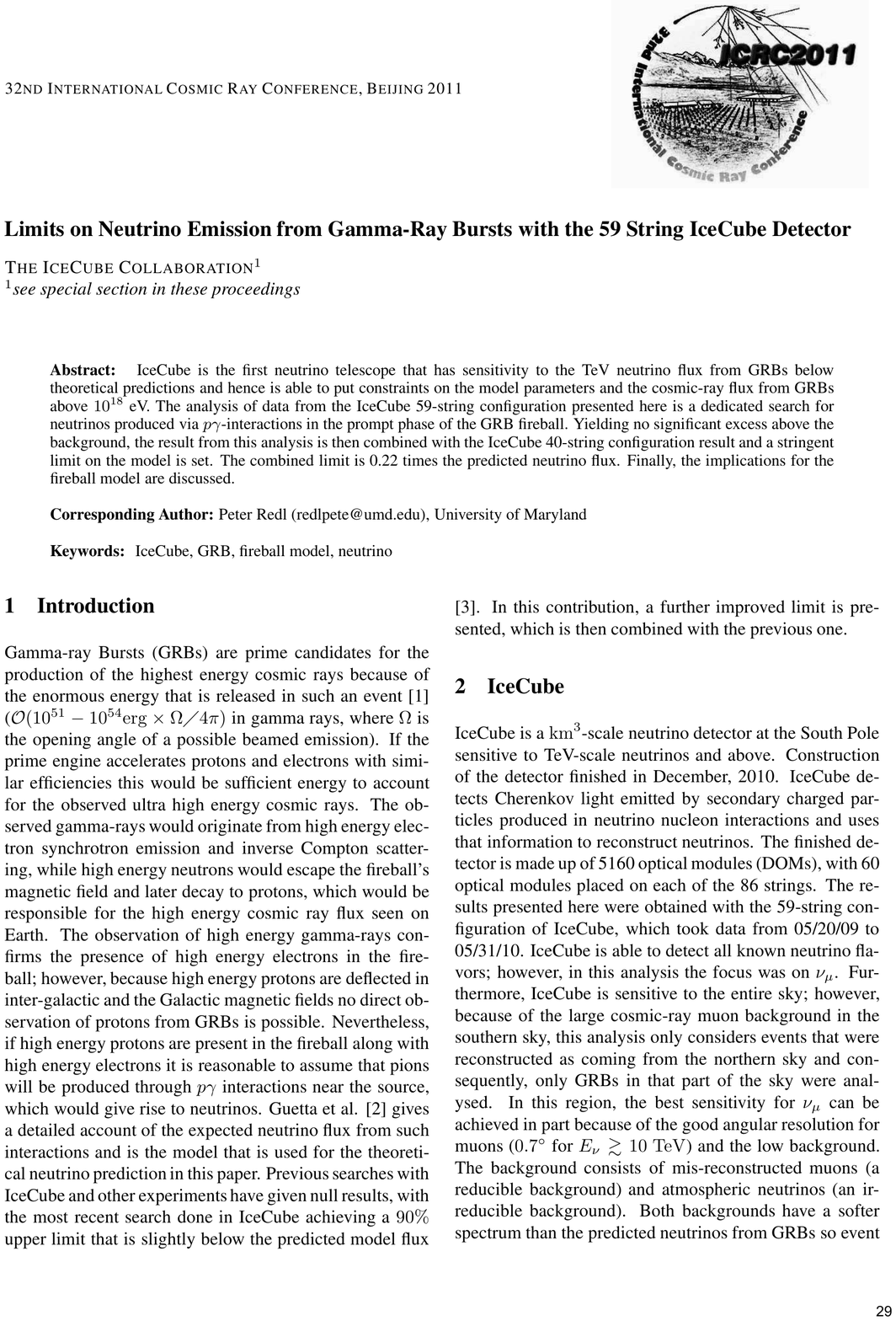}
\end{figure}
\clearpage

\begin{figure}
\includegraphics{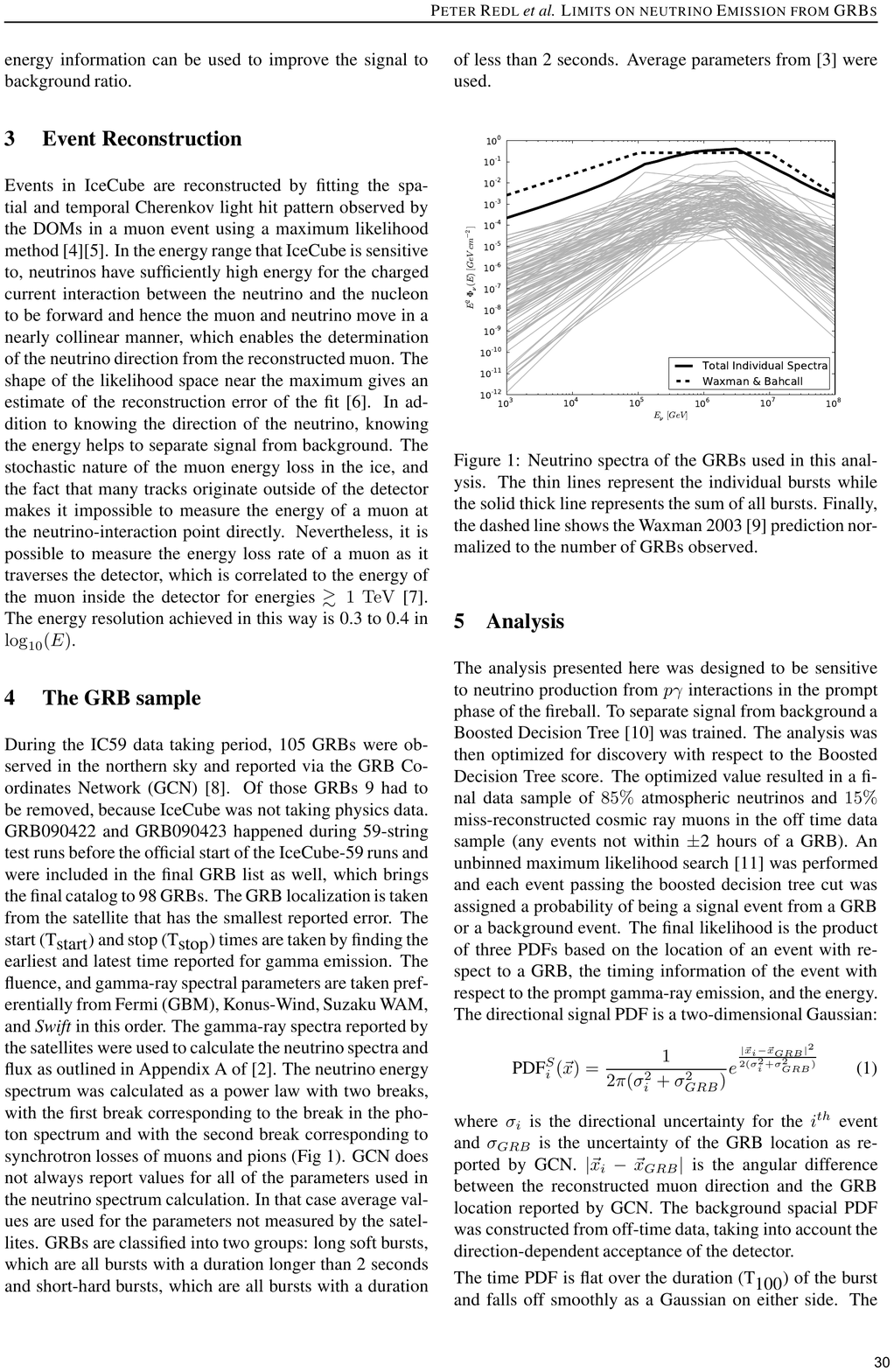}
\end{figure}
\clearpage

\begin{figure}
\includegraphics{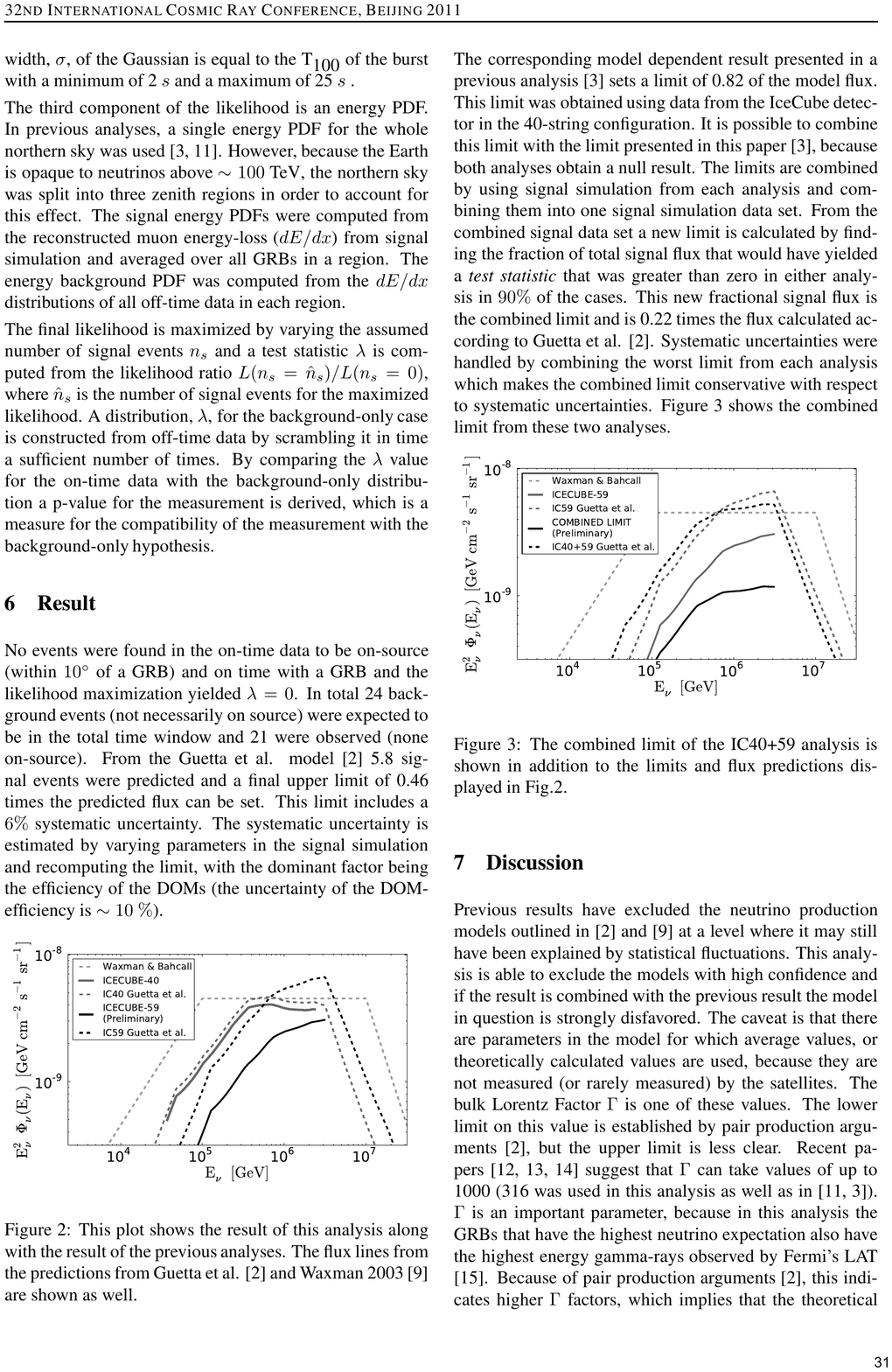}
\end{figure}
\clearpage

\begin{figure}
\includegraphics{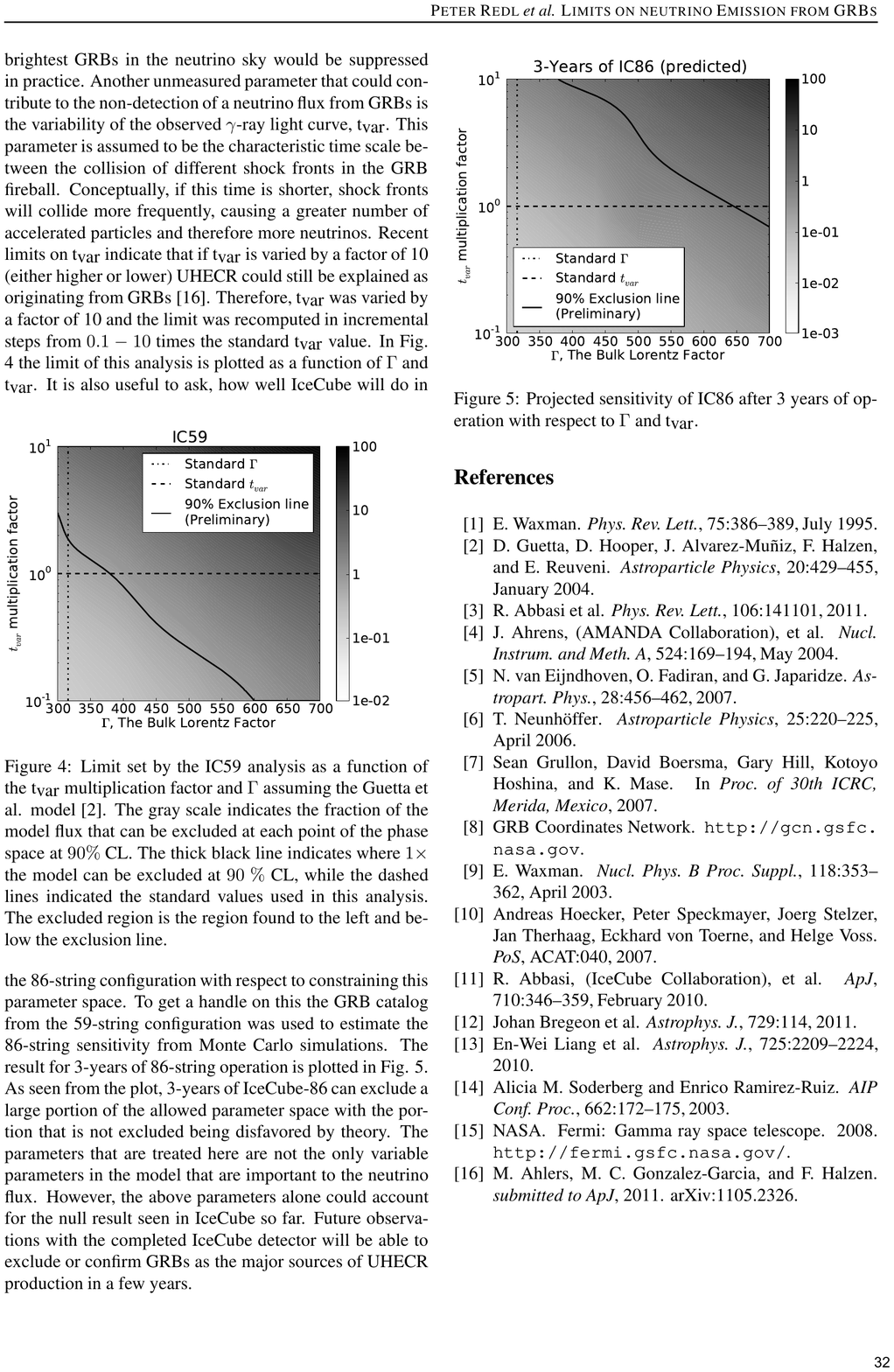}
\end{figure}
\clearpage

\begin{figure}
\includegraphics{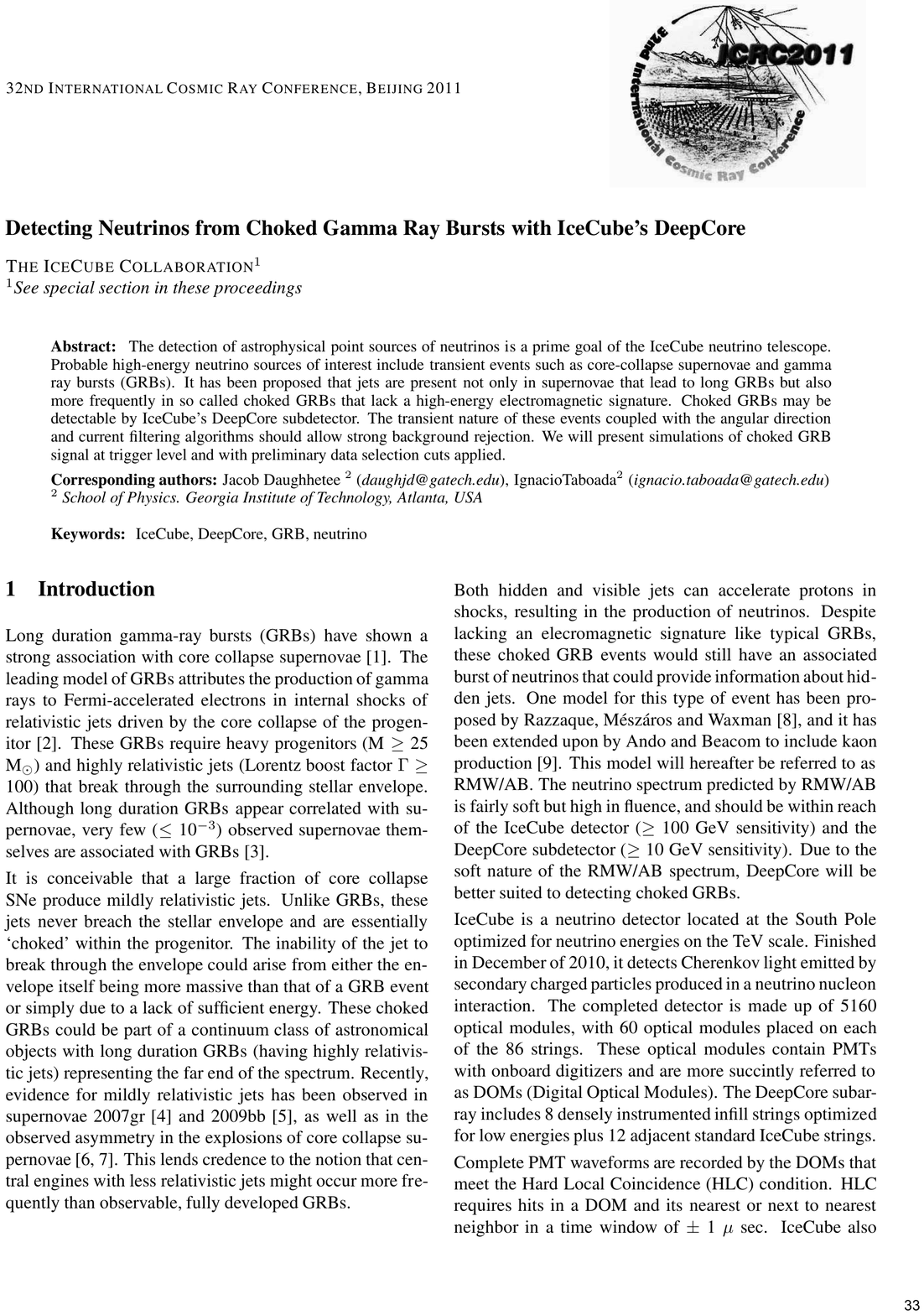}
\end{figure}
\clearpage

\begin{figure}
\includegraphics{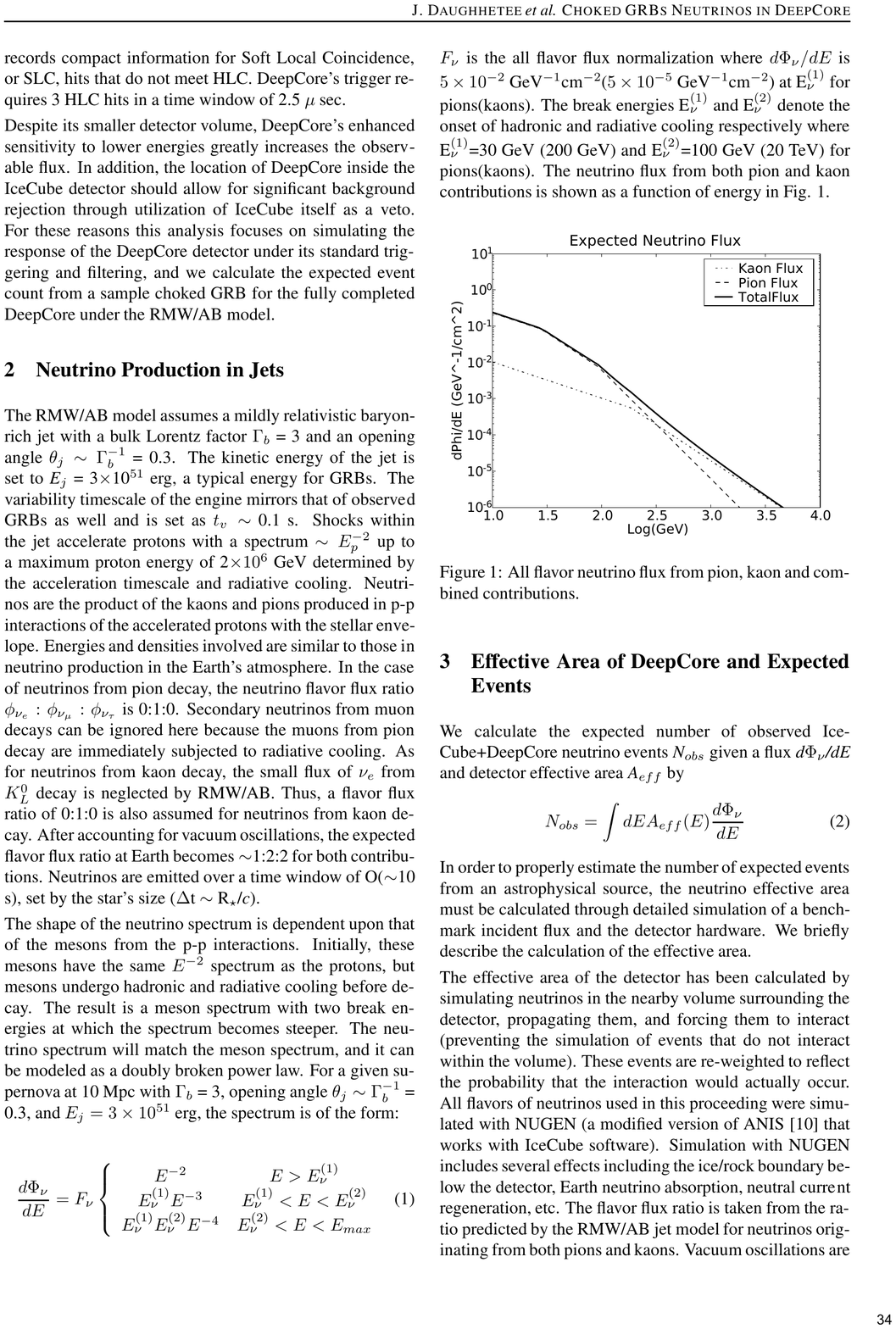}
\end{figure}
\clearpage

\begin{figure}
\includegraphics{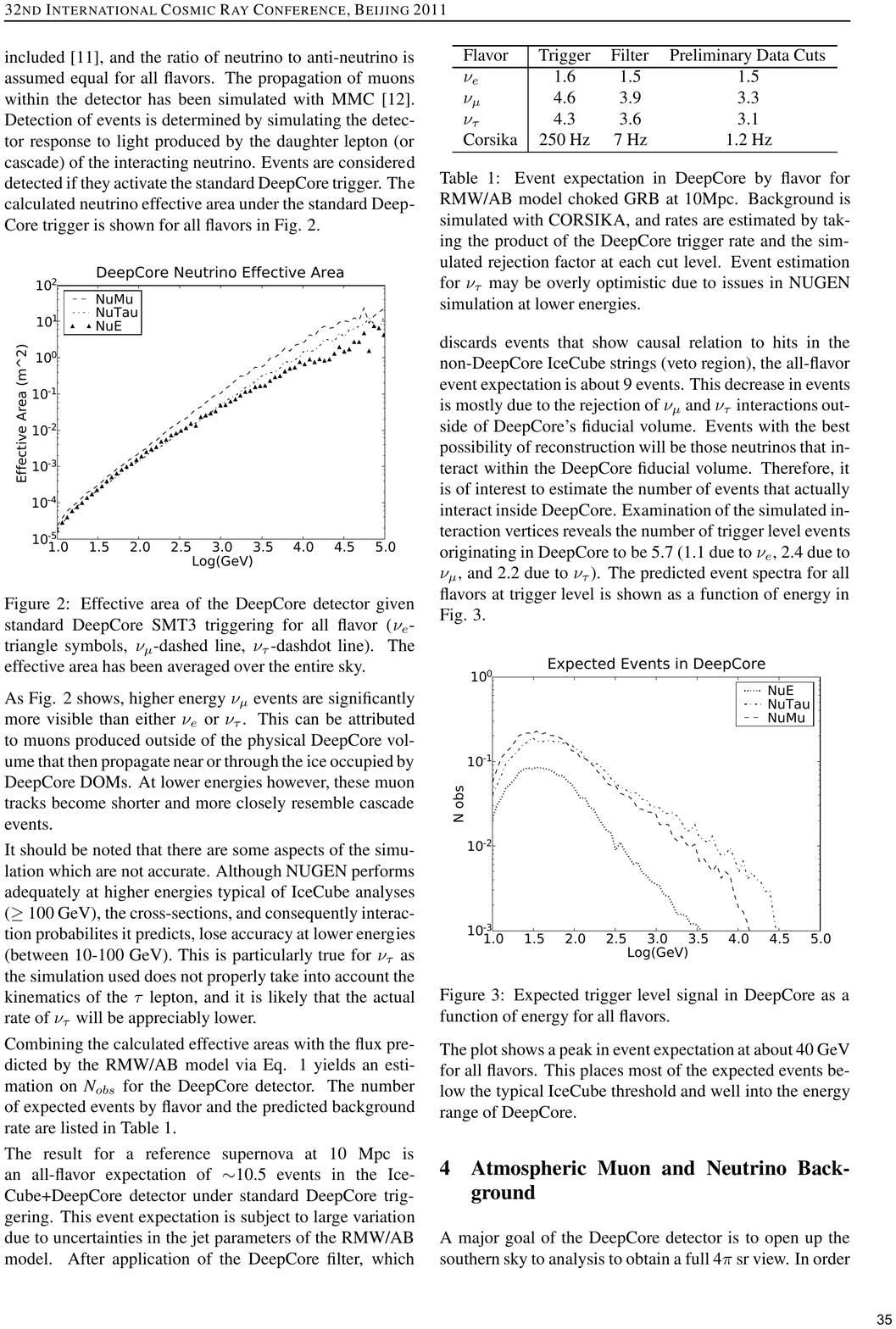}
\end{figure}
\clearpage

\begin{figure}
\includegraphics{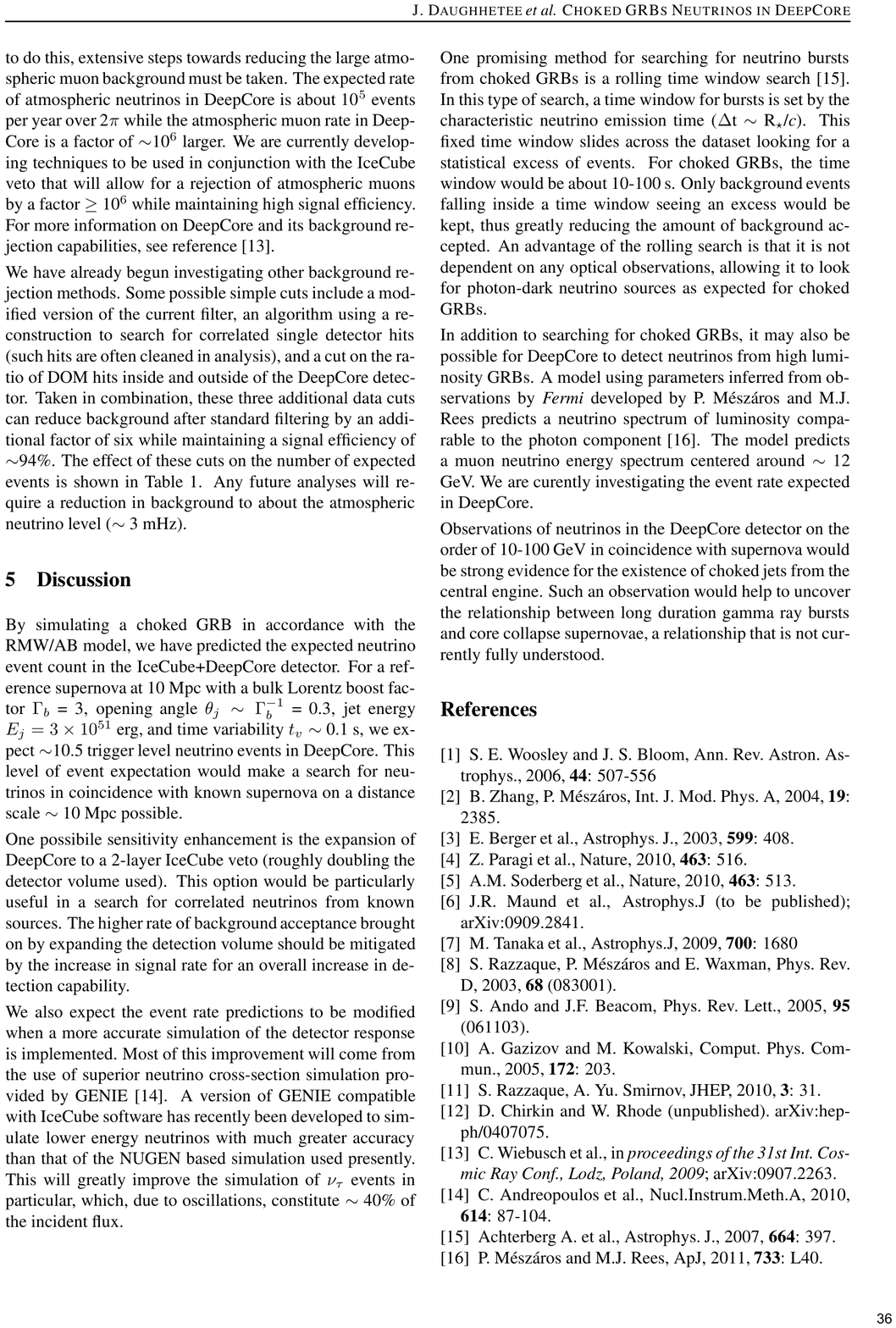}
\end{figure}
\clearpage

\begin{figure}
\includegraphics{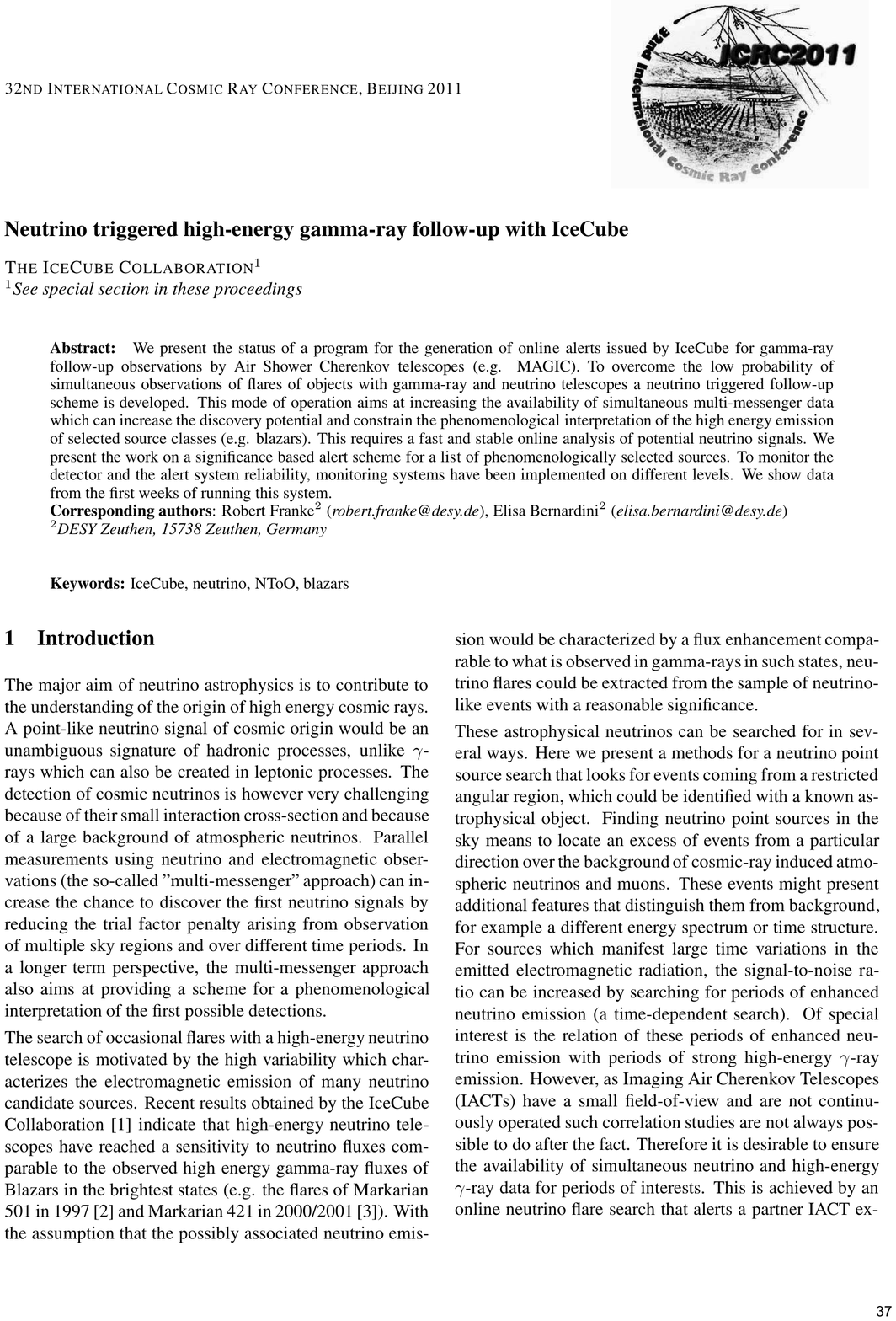}
\end{figure}
\clearpage

\begin{figure}
\includegraphics{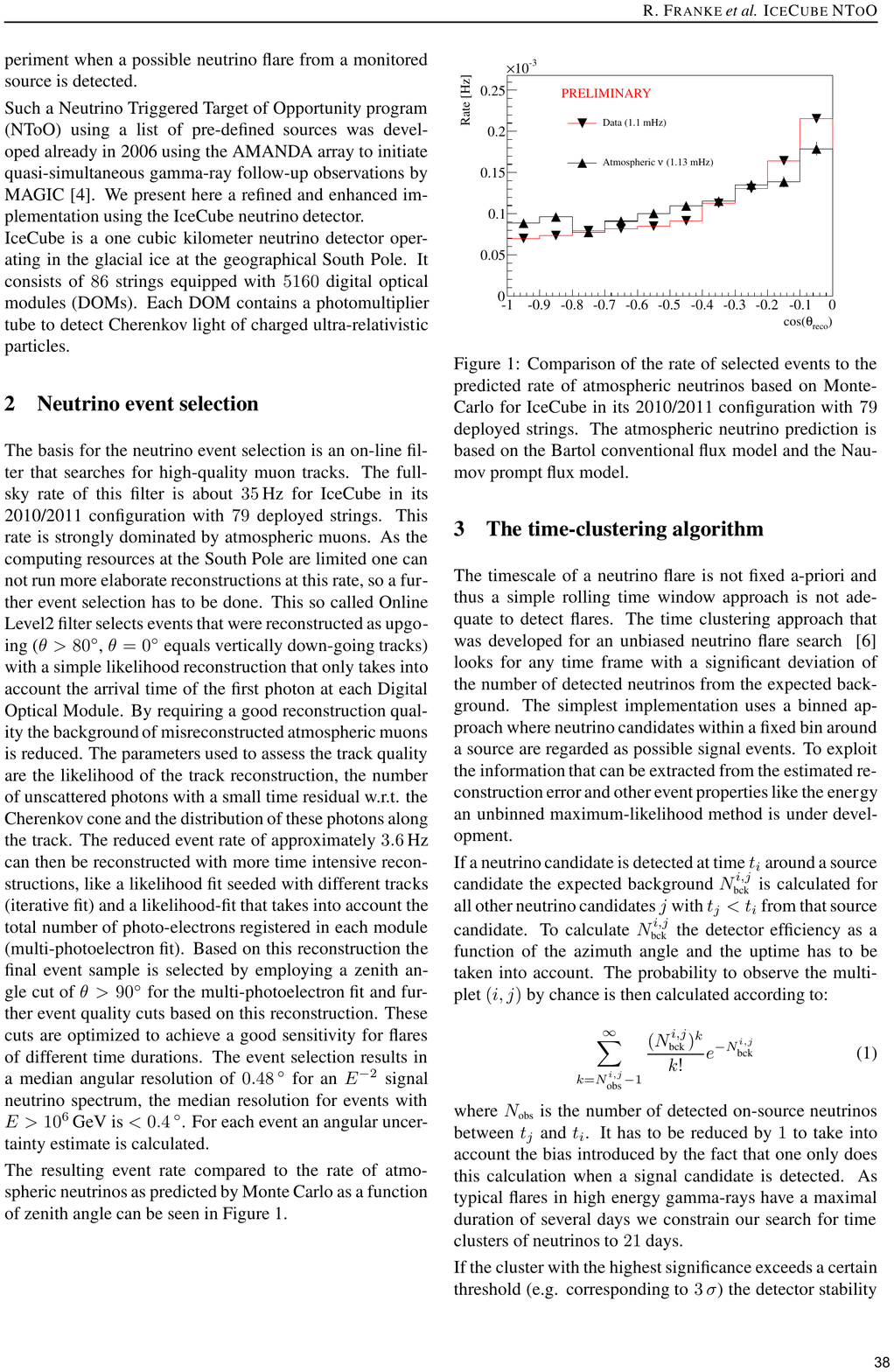}
\end{figure}
\clearpage

\begin{figure}
\includegraphics{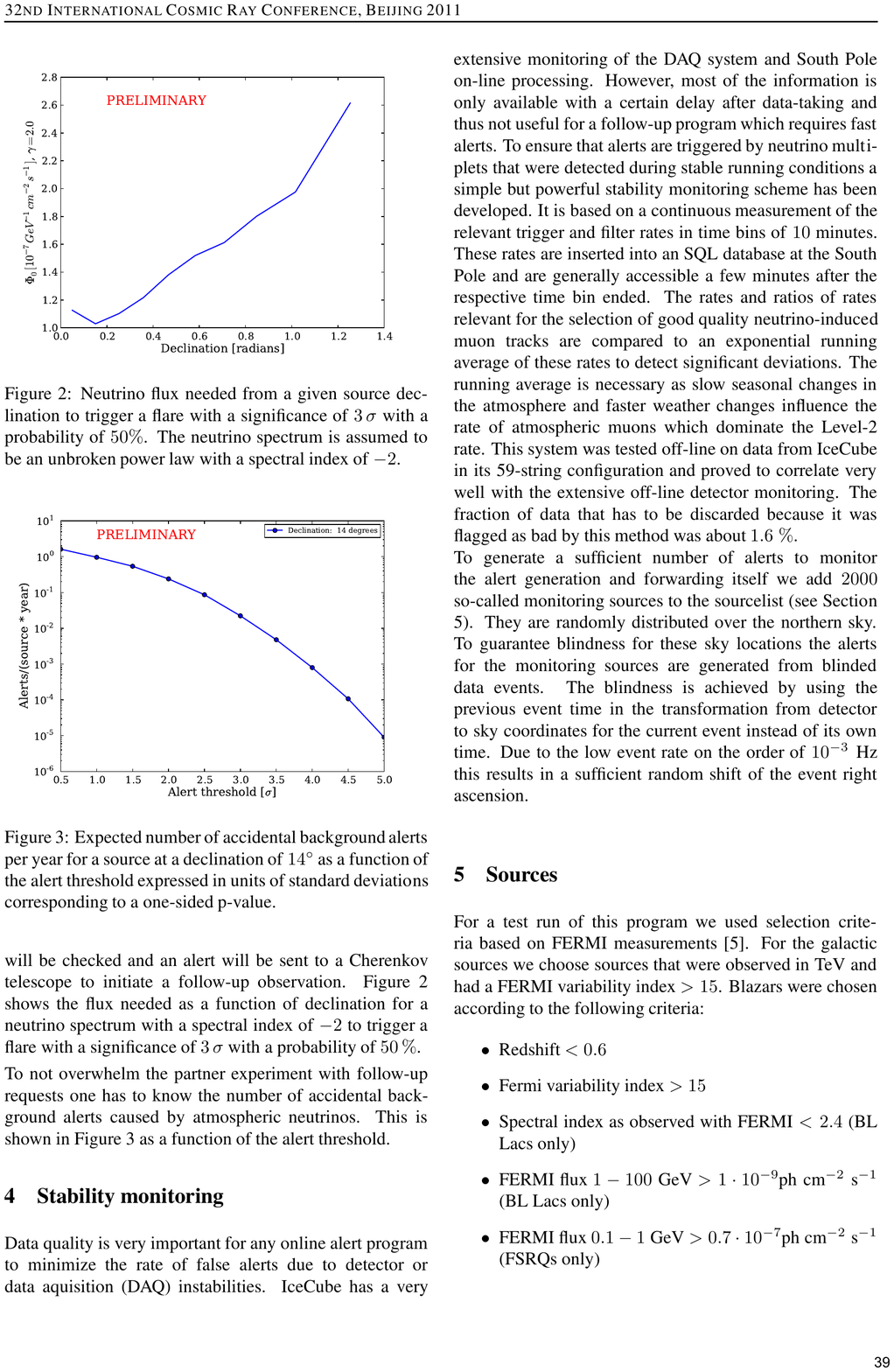}
\end{figure}
\clearpage

\begin{figure}
\includegraphics{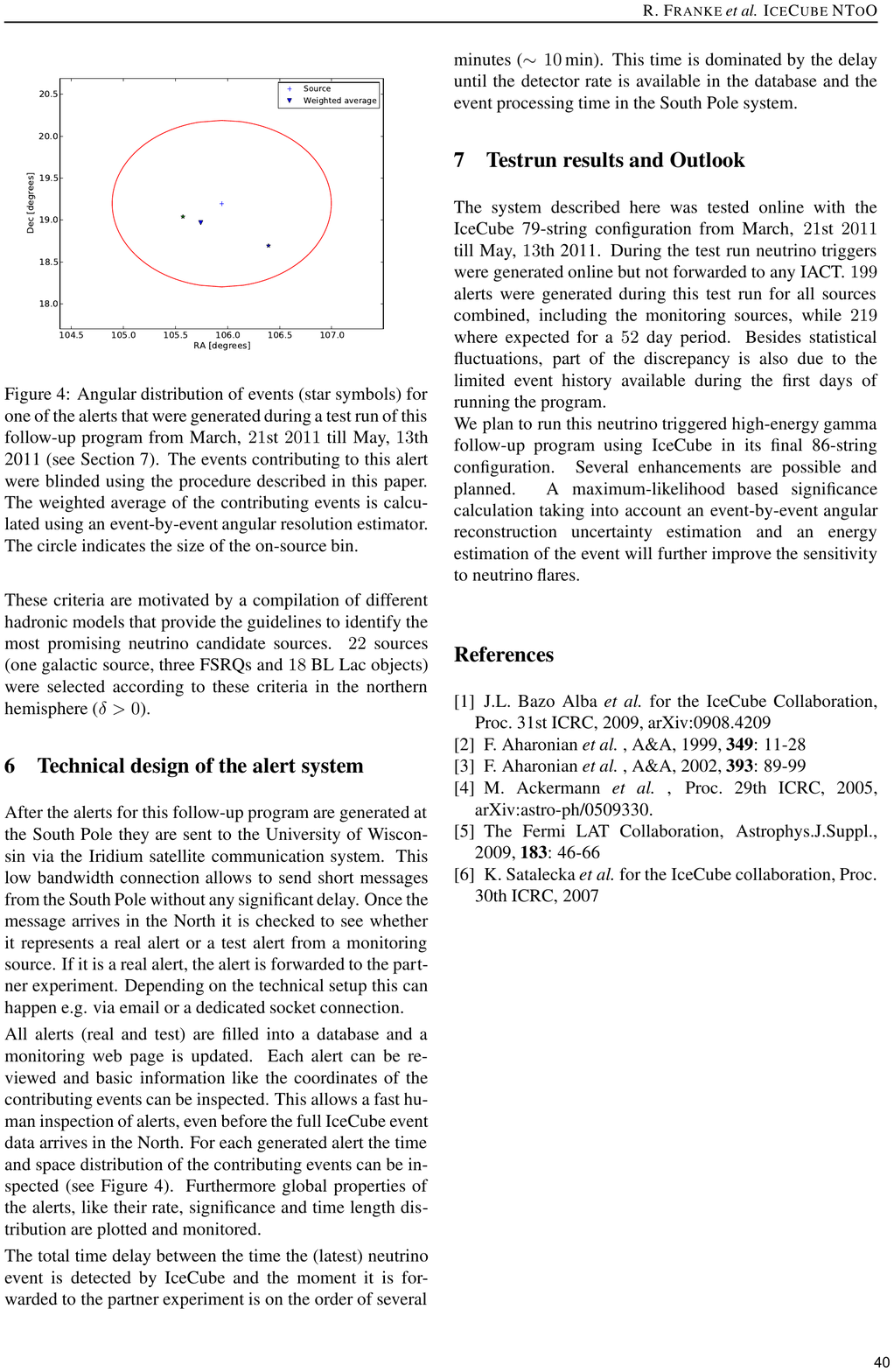}
\end{figure}
\clearpage

\begin{figure}
\includegraphics{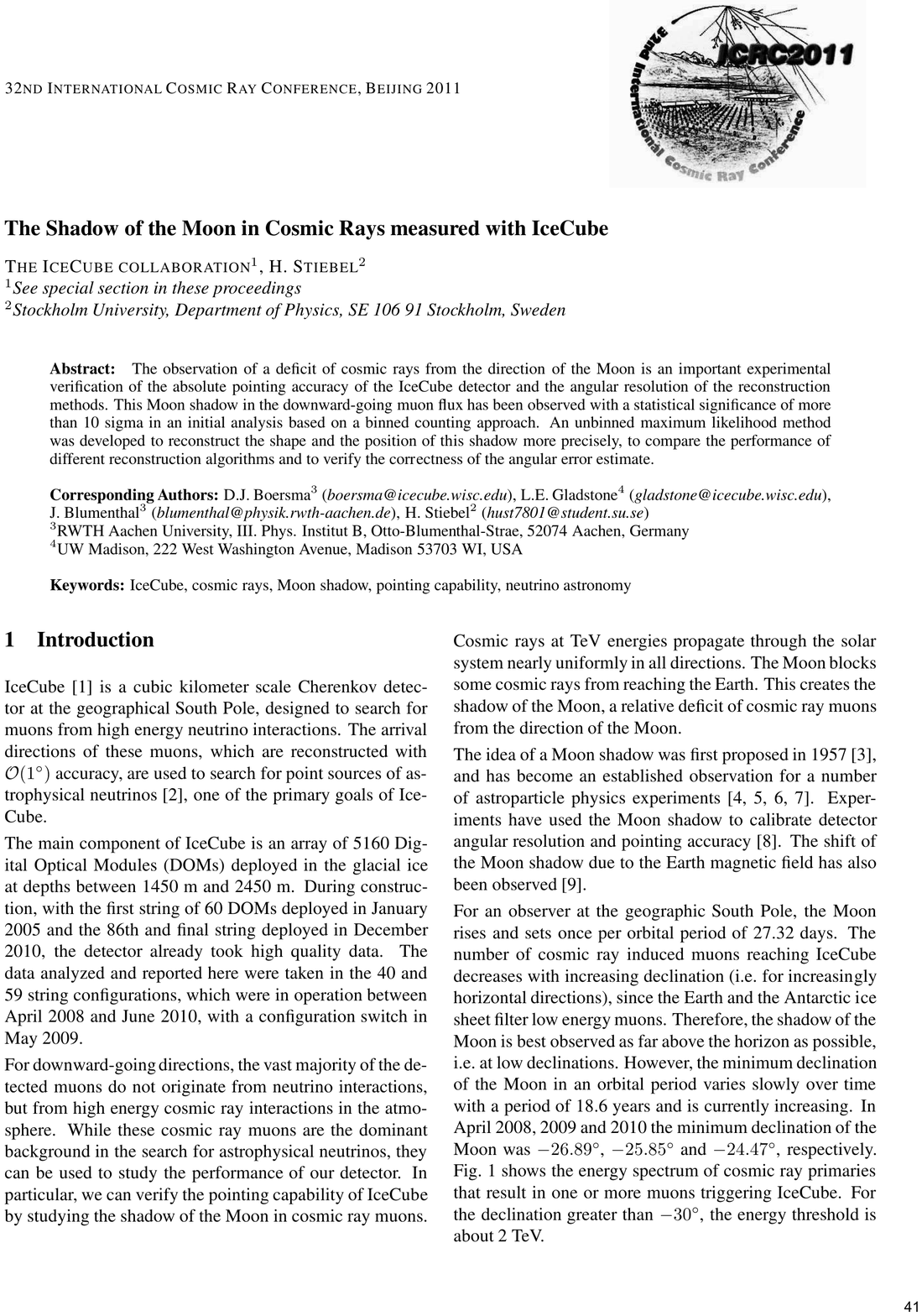}
\end{figure}
\clearpage

\begin{figure}
\includegraphics{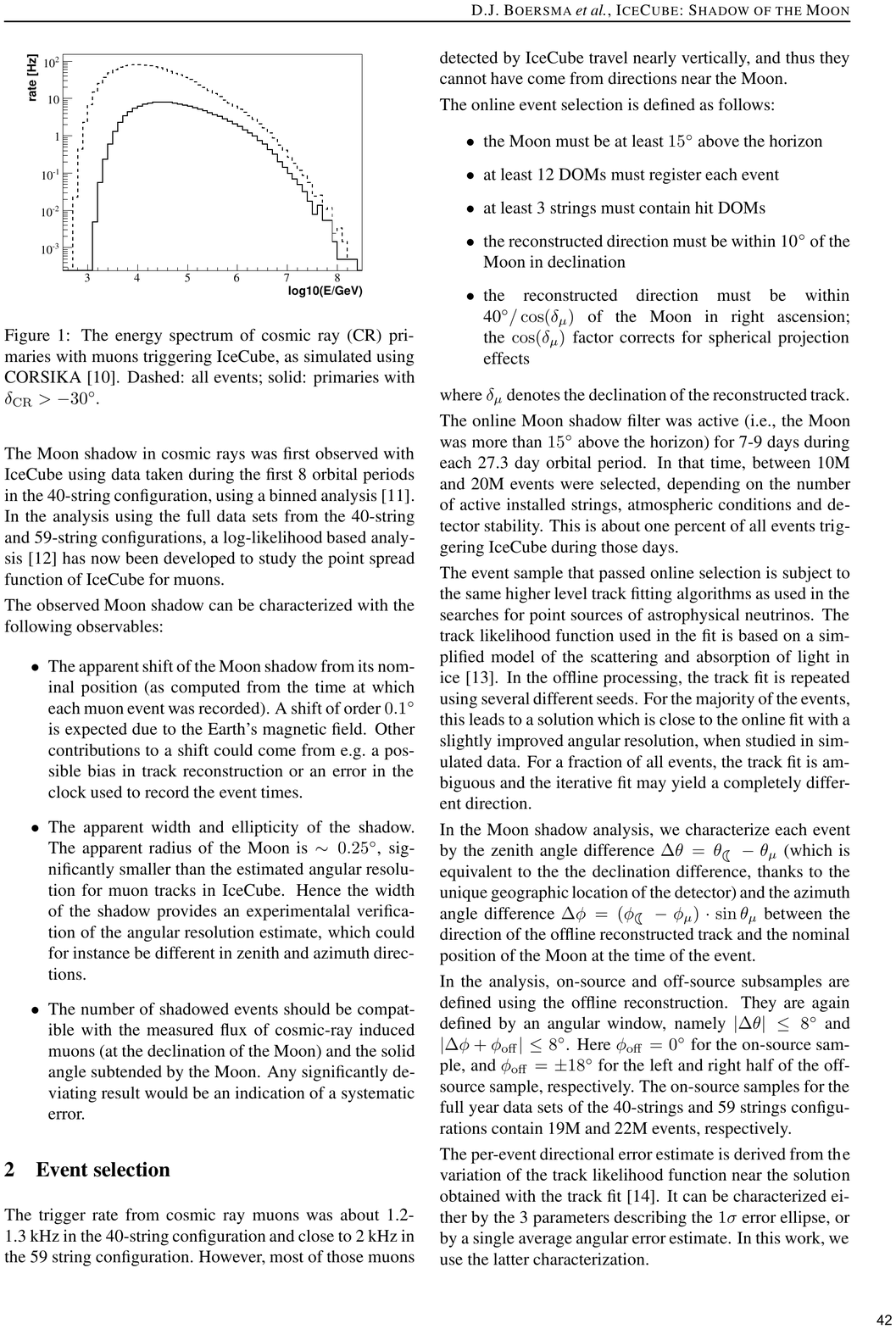}
\end{figure}
\clearpage

\begin{figure}
\includegraphics{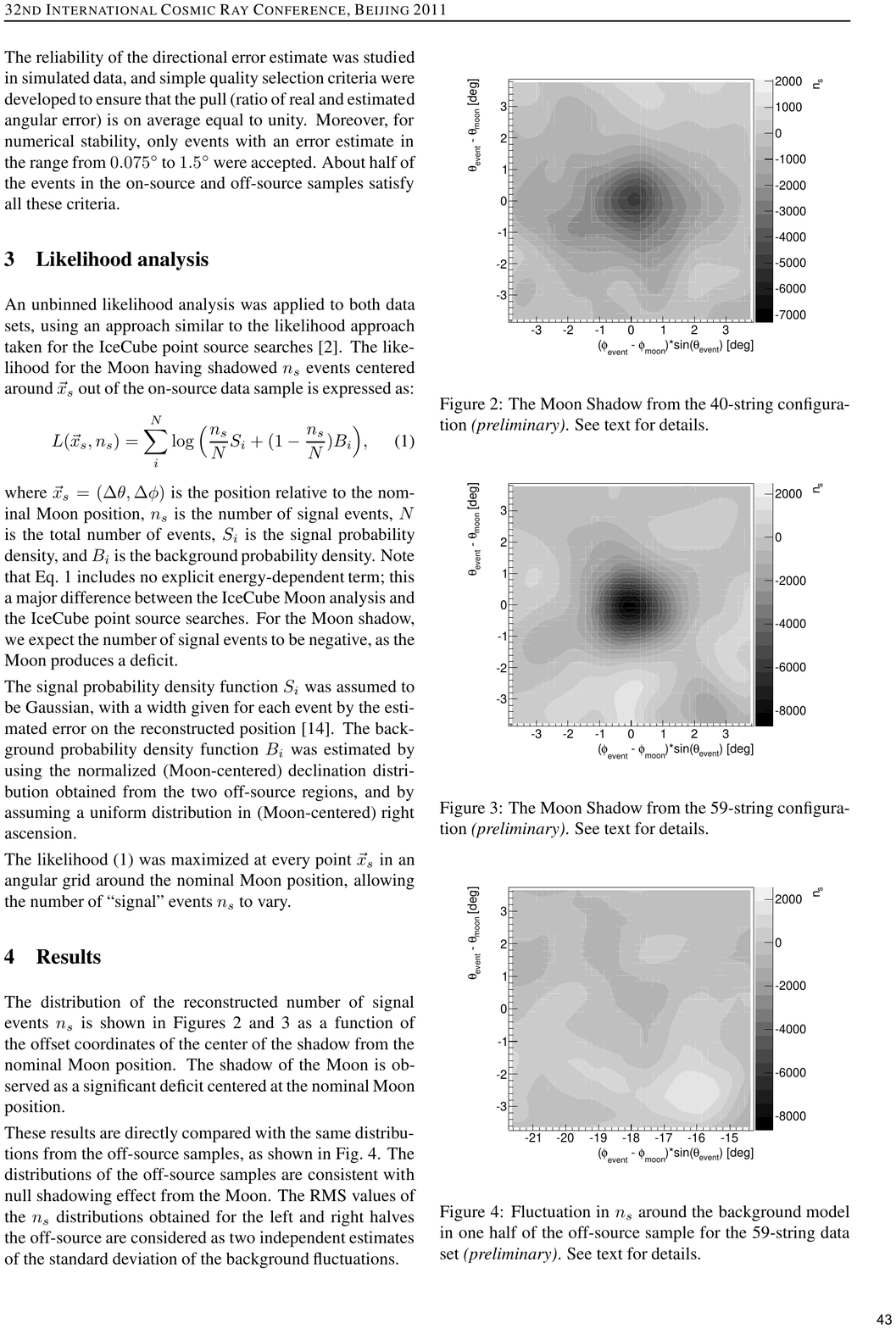}
\end{figure}
\clearpage

\begin{figure}
\includegraphics{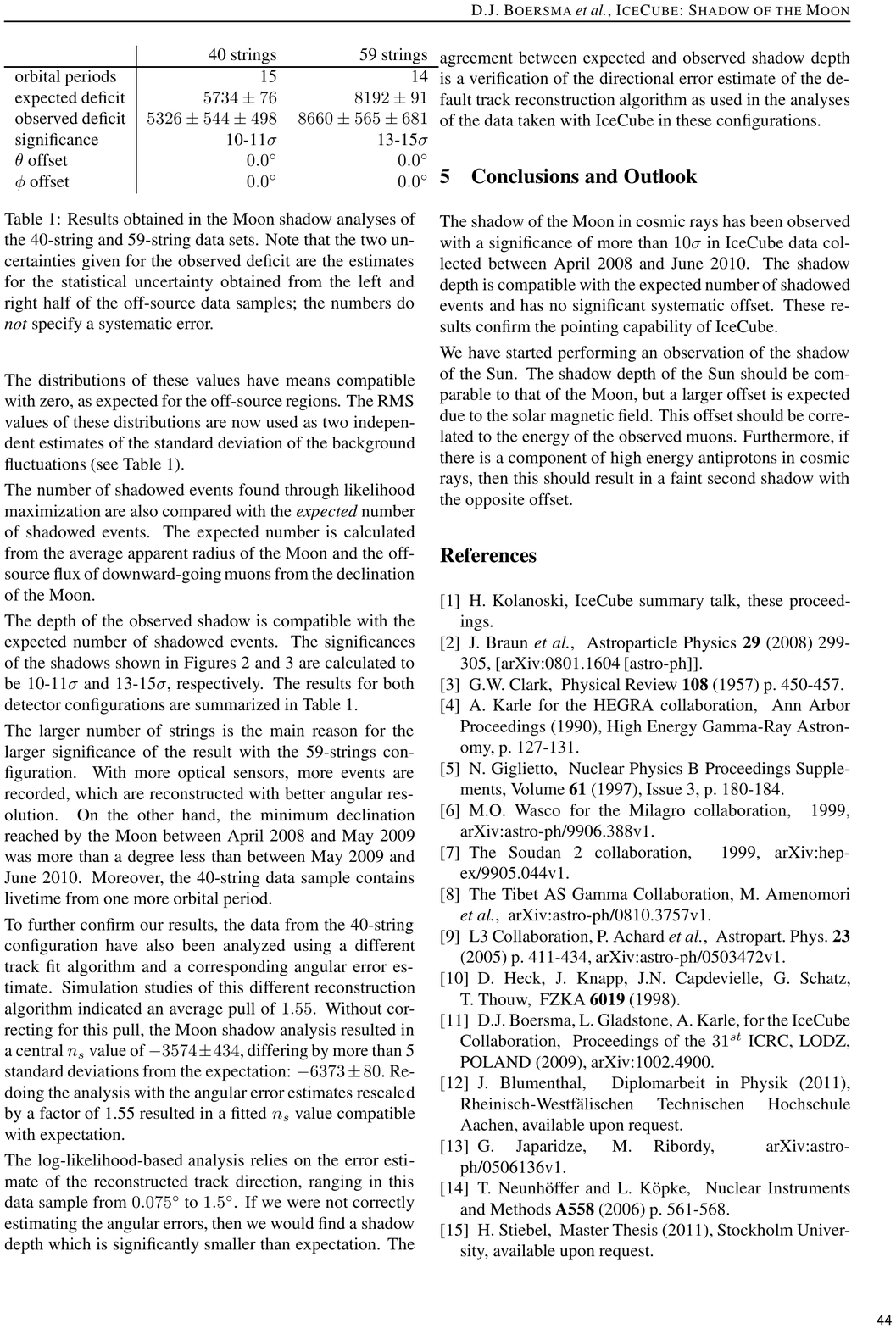}
\end{figure}
\clearpage

\end{document}